\documentclass[twocolumn]{aastex631}

\usepackage{amsmath}
\usepackage{amssymb}
\usepackage{isomath}
\newcommand{\bvec}[1]{\hat{\mathbfit{#1}}}

\usepackage{graphicx}	
\usepackage{bm}
\numberwithin{equation}{section}

\DeclareRobustCommand{\VAN}[3]{#2}
\let\VANthebibliography\thebibliography
\def\thebibliography{\DeclareRobustCommand{\VAN}[3]{##3}\VANthebibliography}

\shorttitle{Fringe-Rate Filtering Signal Loss}
\shortauthors{Pascua et al.}

\begin{document}

\title{A Generalized Method for Characterizing 21-cm Power Spectrum Signal Loss from Temporal Filtering of Drift-scanning Visibilities}

\correspondingauthor{Robert Pascua}
\email{robert.pascua@mail.mcgill.ca}

\author[0000-0003-0073-5528]{Robert Pascua}
\affiliation{
    Department of Physics and Trottier Space Institute, McGill University \\
    3600 University Street, Montreal, QC H3A 2T8, Canada
}
    
\author{Zachary E. Martinot}
\affiliation{
    Department of Physics and Astronomy, University of Pennsylvania \\
    209 S 33rd St, Philadelphia, PA 19104, USA
}

\author[0000-0001-6876-0928]{Adrian Liu}
\affiliation{
    Department of Physics and Trottier Space Institute, McGill University \\
    3600 University Street, Montreal, QC H3A 2T8, Canada
}
    
\author[0000-0002-4810-666X]{James E. Aguirre}
\affiliation{
    Department of Physics and Astronomy, University of Pennsylvania \\
    209 S 33rd St, Philadelphia, PA 19104, USA
}

\author[0000-0002-8211-1892]{Nicholas S. Kern}
\altaffiliation{NASA Hubble Fellow}
\affiliation{
    Department of Physics, Massachusetts Institute of Technology \\ 
    182 Memorial Dr, Cambridge, MA 02139, USA
}

\author[0000-0003-3336-9958]{Joshua S. Dillon}
\affiliation{
    Department of Astronomy and Radio Astronomy Laboratory, University of California, Berkeley \\
    501 Campbell Hall, Berkeley, CA 94720, USA
}

\author[0000-0001-7716-9312]{Michael J. Wilensky}
\altaffiliation{CITA National Fellow}
\affiliation{
    Department of Physics and Trottier Space Institute, McGill University \\
    3600 University Street, Montreal, QC H3A 2T8, Canada
}

\author[0000-0001-5300-3166]{Nicolas Fagnoni}
\affiliation{
    Cavendish Astrophysics, University of Cambridge \\
    JJ Thomson Avenue, Cambridge CB3 0HE, UK
}

\author[0000-0001-8530-6989]{Eloy de Lera Acedo}
\affiliation{
    Cavendish Astrophysics, University of Cambridge \\
    JJ Thomson Avenue, Cambridge CB3 0HE, UK
}
\affiliation{
    Kavli Institute for Cosmology in Cambridge, University of Cambridge \\
    Madingley Road, Cambridge CB3 0EZ, UK
}

\author[0000-0003-3197-2294]{David R. DeBoer}
\affiliation{
    Department of Astronomy and Radio Astronomy Laboratory, University of California, Berkeley \\
    501 Campbell Hall, Berkeley, CA 94720, USA
}



\begin{abstract}
A successful detection of the cosmological 21-cm signal from intensity mapping experiments (for example, during the Epoch of Reioinization or Cosmic Dawn) is contingent on the suppression of subtle systematic effects in the data.
Some of these systematic effects, with mutual coupling a major concern in interferometric data, manifest with temporal variability distinct from that of the cosmological signal.
Fringe-rate filtering---a time-based Fourier filtering technique---is a powerful tool for mitigating these effects; however, fringe-rate filters also attenuate the cosmological signal.
Analyses that employ fringe-rate filters must therefore be supplemented by careful accounting of the signal loss incurred by the filters.
In this paper, we present a generalized formalism for characterizing how the cosmological 21-cm signal is attenuated by linear time-based filters applied to interferometric visibilities from drift-scanning telescopes.
Our formalism primarily relies on analytic calculations and therefore has a greatly reduced computational cost relative to traditional Monte Carlo signal loss analyses.
We apply our signal loss formalism to a filtering strategy used by the Hydrogen Epoch of Reionization Array (HERA) and compare our analytic predictions against signal loss estimates obtained through a Monte Carlo analysis.
We find excellent agreement between the analytic predictions and Monte Carlo estimates and therefore conclude that HERA, as well as any other drift-scanning interferometric experiment, should use our signal loss formalism when applying linear, time-based filters to the visibilities.
\end{abstract}



\section{Introduction}
Many of the current generation of low-frequency radio interferometers have made great strides in constraining the power spectrum of the cosmological 21-cm signal at high redshift~\citep[e.g.,][]{Li:2019,Yoshiura:2021,HERA:2022b,HERA:2023,LOFAR:2020,Kolopanis:2023,NenuFAR:2024}, and it is anticipated that future observations with the current generation and next generation of experiments will provide the first detection of the cosmological signal at high redshift~\citep{Koopmans:2015,Braun:2019,Mertens:2021,Breitman:2023}.
These constraints are supplemented by multiple low-redshift ($z < 1$) detections of the cosmological 21-cm signal in cross-correlation~\citep{Chang:2010,Masui:2013,Anderson:2018,Cunnington:2022,CHIME:2023a,CHIME:2023b} and one claimed detection in auto-correlation~\citep{Paul:2023}.
The constraints obtained from these experiments have already begun to rule out exotic reioinization scenarios, with the most stringent constraints ruling out a broad class of ``cold reionization'' scenarios~\citep{Greig:2020,Ghara:2021,HERA:2022a,HERA:2023}.
As low-frequency radio cosmology experiments push to deeper limits and an eventual detection, we will obtain unique insights to the physics of Cosmic Dawn (CD) and the Epoch of Reionization (EoR)~\citep[e.g.,][]{Madau:1997,Furlanetto:2006,Pritchard&Loeb:2012,Liu&Shaw:2020}.
A successful detection, however, is contingent on extremely high sensitivity and a high degree of control over systematics in the data, since the cosmological 21-cm signal is expected to be roughly $10^5$ times fainter than the synchrotron-dominated foreground signal~\citep{Liu&Shaw:2020}.

Experiments typically take one of two approaches to obtain the extreme sensitivity required for detecting the high-redshift cosmological 21-cm signal.
Phased arrays like the Murchison Widefield Array~\citep[MWA,][]{Tingay:2013,Bowman:2013}, the LOw-Frequency ARray~\citep[LOFAR,][]{vanHaarlem:2013}, NenuFAR~\citep{Zarka:2018}, and the Square Kilometre Array~\citep[SKA,][]{Koopmans:2015} typically build up sensitivity by phasing to a few small, cold patches of the sky and obtaining very deep observations of these patches.
While the MWA computes power spectra directly from the visibilities~\citep[e.g.,][]{Trott:2016,Yoshiura:2021,Kolopanis:2023} as well as from image cubes~\citep[e.g.,][]{Jacobs:2016,Li:2019}, LOFAR and NenuFAR opt to compute power spectra strictly from images~\citep[e.g.,][]{LOFAR:2020,NenuFAR:2024} and the SKA intends to follow suit and eventually tomographically map the cosmological 21-cm signal~\citep{Koopmans:2015,Braun:2019}.
Drift scan arrays like the Hydrogen Epoch of Reionization Array~\citep[HERA,][]{DeBoer:2017,Berkhout:2024}, the Canadian Hydrogen Observatory and Radio-transient Detector~\citep[CHORD,][]{Vanderlinde:2020}, the Precision Array for Probing the Epoch of Reionization~\citep[PAPER,][]{Parsons:2012a}, and the Hydrogen Intensity Real-time Analysis eXperiment~\citep[HIRAX,][]{Newburgh:2016,Chrichton:2022} instead build up the requisite sensitivity through instantaneous \emph{redundancy}~\citep{Dillon&Parsons:2016}.
This strategy requires an array of identical antennas to be placed on a regular grid, which enables the collection of many identical visibility measurements with independent noise realizations, since the interferometric response is uniquely characterized by the physical separation, or baseline, between the pair of antennas used to form the visibility.
In this case, sensitivity is built up through averaging ``redundant'' visibilities together, with a further increase in sensitivity obtained by coherently averaging together observations of the same patch of sky across many nights of observing (i.e., these experiments primarily build up sensitivity through extensive averaging in the $uv$-plane).
In short, phased arrays with very little instantaneous redundancy (like the MWA, LOFAR, and SKA) build up sensitivity through repeated deep observations of small patches of sky, while redundant drift-scanning arrays (like HERA, CHORD, and HIRAX) quickly accumulate sensitivity over larger patches of sky at the cost of reduced imaging capabilities.

Although both of these strategies are viable for accumulating the sensitivity required to detect the cosmological 21-cm signal, they are both prone to a variety of systematic effects, as evidenced by the reported excess power above the expected thermal noise floor in some cosmological Fourier modes across a broad range of redshifts~\citep[e.g.,][]{LOFAR:2020,HERA:2022b,HERA:2023,Kolopanis:2023,NenuFAR:2024}.
Potential sources of these systematic effects include Radio Frequency Interference (RFI) from anthropogenic sources~\citep[e.g.,][]{Wilensky:2023b}, calibration and beam modelling errors~\citep[e.g.,][]{Barry:2016,Ewall-Wice:2017,Byrne:2019,Orosz:2019,Barry&Chokshi:2022}, ionospheric effects~\citep[e.g.,][]{Martinot:2018}, and mutual coupling between antennas~\citep[e.g.,][]{Kern:2019,Kern:2020a,Josaitis:2022,Rath&Pascua:2024}.

Mutual coupling is a particularly pressing issue that needs to be addressed.
In the context of phased arrays,~\citet{Bolli:2022} showed that the presence of just one nearby antenna produces strong spatial and spectral distortions in the direction-dependent antenna response (colloquially, the ``beam'').
This poses serious challenges for precision direction-dependent calibration and foreground removal techniques, and could potentially remove the possibility of detecting the cosmological 21-cm signal if not properly dealt with~\citep{Byrne:2019,Barry&Chokshi:2022}.
In the context of drift scan arrays, \citet{Rath&Pascua:2024} showed that simulations based on the re-radiative mutual coupling model described in~\citet{Josaitis:2022} produced systematic features in the visibilities that were broadly consistent with systematic features seen in HERA data, and that these features dominated over a plausible cosmological signal across a wide range of cosmological Fourier modes.
In addition to this, \citet{Rath&Pascua:2024} built on the work of~\citet{Parsons:2016} and~\citet{Kern:2019} by further investigating the utility of fringe-rate filtering as a mutual coupling mitigation tool, finding that simple fringe-rate filters could mitigate roughly 90\% of the coupling features in the visibilities (or roughly 99\% of the excess in the power spectrum).

While~\citet{Rath&Pascua:2024} showed that fringe-rate filtering reduces contamination from mutual coupling, the signal loss properties of such filtering schemes were left to be investigated here.
Signal loss analyses are absolutely crucial when proposing new analysis techniques, since failing to correctly account for the signal attenuation incurred by the filter can result in artificially low power spectrum estimates that misrepresent the true power spectrum estimates~\citep[see e.g.,][]{Switzer:2015,Cheng:2018}.
In other words, since fringe-rate filtering can attenuate the cosmological signal, power spectrum estimates made with filtered data must use a different normalization factor than estimates made with unfiltered data, and signal loss analyses can provide the required additional normalization.

In addition to extending the work of~\citet{Parsons&Backer:2009} by exploring how fringe-rate filtering can be leveraged to minimize the variance in power spectrum estimates and minimize polarization leakage,~\citet{Parsons:2016} also computed a modified power spectrum normalization that takes into account the effects of fringe-rate filtering.
\citet{Parsons:2016} found that the power spectrum normalization can be computed in the usual way (i.e., as in~\citealt{Parsons:2014}), but with the beam integral terms computed using an ``effective'' primary beam.
Their analysis, however, crucially depended on the assumption that the temporal evolution of the visibilities was dominated by sources drifting through the interferometric fringe pattern, and that any time variability from sources drifting through the primary beam was negligible.
This approximation begins to break down for baselines that are only a few dish diameters long, since the fringe period is comparable to the beam crossing time in this limit (i.e., when a source transits the main lobe of the beam, the interferometric delay changes on roughly the same timescales that the transmission amplitude changes).
Since close-packed arrays are ubiquitous among the current and next generation of drift scan radio cosmology experiments, it is both timely and important to develop a signal loss formalism that extends beyond the formalism of~\citet{Parsons:2016}.

A key strategy in our rigorous treatment of fringe-rate filtering is the use of the $m$-mode formalism~\citep{Shaw:2014}.
In a coordinate system where the polar axis is aligned with the Earth's spin axis, the azimuthal spherical harmonic index $m$ is the Fourier dual to the azimuthal angle.
For drift scan observations, the telescope is pointed at an azimuthal angle that grows linearly with time, and therefore the $m$ index can also be thought of as indexing the Fourier transform of a full sidereal day of visibility measurements.
Fringe-rate space is accessed by Fourier transforming a time stream of visibilities (though not necessarily a full sidereal day's worth of them---a crucial point that we will return to in Section~\ref{sec:FRINGE-RATE-COVARIANCE}), so a detailed understanding of fringe-rate filtering can be achieved by using the machinery of the $m$-mode formalism.
In this paper, we provide exactly this detailed understanding by building on complementary works in the literature.
We draw on analytic tools from the $m$-mode formalism as developed in Martinot \& Aguirre (in preparation) and reviewed in this paper that provides additional tools for understanding the simulation-based results of~\citet{Garsden:2024}.
We use a suite of Monte Carlo simulations to validate our formalism's analytic prediction for signal loss due to fringe-rate filtering.
Of course, such a validation exercise is only possible in the context of a statistically disciplined framework for making comparisons, which we provide.
Our goal in this paper is to focus on a thorough investigation of signal loss, but harmonic analyses of drift-scan radio telescopes have much broader applicability beyond signal loss calculations.
For example, as shown in Martinot \& Aguirre (in preparation), harmonic analyses of drift-scan radio telescopes may be used to derive optimal signal-to-noise maximizing Wiener filters and to critically examine the regimes in which such filters reduce to commonly used techniques such as fringe-stopped averaging.

The remainder of this paper is organized as follows: In Section~\ref{sec:TIME-EVOLUTION}, we review the more general framework of Martinot \& Aguirre (in preparation) for characterizing the time-harmonic response of a drift-scanning interferometer; In Section~\ref{sec:SIGNAL-LOSS}, we describe the procedure for calculating the expected signal loss associated with a given fringe-rate filter; In Section~\ref{sec:SIMULATIONS}, we compare our signal loss methodology against numerical simulations; In Section~\ref{sec:CONCLUSION} we summarize our findings.

\section{Time Evolution of Visibilities}
\label{sec:TIME-EVOLUTION}
Time-based filtering operations, such as fringe-rate filtering and coherent time averaging, are frequently employed in the analysis of interferometric data from drift scan arrays.
These operations typically attenuate the cosmological signal and must therefore be accompanied by a disciplined accounting of the associated signal loss.
Typically, the signal loss is estimated numerically~\citep[e.g.,][]{Cheng:2018,Aguirre:2022} or analytically treated using approximate methods~\citep[e.g.,][]{Parsons:2016}---an \emph{exact} treatment of how to analytically characterize the signal loss, however, has not yet been proposed in the literature.

Characterizing the time-time covariance is an essential first step in developing a statistically disciplined method for computing the signal loss from a time-based filter.
This is a consequence of three key points:
First, the average measured power is determined by the variance in the visibilities, which is just the diagonal of the time-time covariance;
Second, since the instrument response induces correlations on the sky, drift-scan observations have non-trivial correlations in time (i.e., non-diagonal structure in the time-time covariance);
Third, time-based filters universally combine data from different times, which mixes diagonal and off-diagonal modes in the time-time covariance.
Taken together, these key ideas mean that a statistically disciplined approach for computing the filter-induced signal loss must begin with the time-time covariance.

In this section we review the framework of Martinot \& Aguirre (in preparation) for characterizing the time-time covariance of the interferometric visibility $V(t)$,
\begin{equation}
    \label{eq:time-time-covariance-definition}
    C(t,t') \equiv \bigl\langle V(t) V(t')^* \bigr\rangle - \bigl\langle V(t) \bigr\rangle \bigl\langle V(t')^* \bigr\rangle,
\end{equation}
where the angled brackets $\langle \cdot \rangle$ indicate an ensemble average.
The framework we review requires a model of the antenna's primary beam (which may be analytic, modeled via computational electromagnetism simulations, or measured empirically with drones or bright source transits, for example), but all other calculations can be performed analytically.
In the remainder of this section, we provide a derivation of this framework, discuss how to extend it to other bases, and compare our results against those of~\citet{Parsons:2016}.

\subsection{Characterizing the Covariance}
\label{sec:TIME-TIME-COVARIANCE}
We aim to compute the time-time covariance in the visibilities measured by a drift-scan telescope, under the assumption that the Earth's rotational rate is fixed and the orientation of its rotational axis is fixed relative to a celestial frame.
In other words, we are ignoring various kinematic effects like precession and nutation, and we are assuming that the measured visibilities are truly periodic over one sidereal day.
Additionally, we restrict our attention to unpolarized visibilities and work in an Earth-centered, Earth-fixed coordinate system, so the measured visibility at time $t$ is
\begin{equation}
    \label{eq:measurement-equation}
    V(t) = \int_{4\pi} A(\hat{\mathbfit{n}}, t) I(\hat{\mathbfit{n}}) e^{-i2\pi\nu\mathbfit{b}(t) \cdot \hat{\mathbfit{n}}/c} d\Omega,
\end{equation}
where $A(\hat{\mathbfit{n}},t)$ is the peak-normalized primary beam, $I(\hat{\mathbfit{n}})$ is the specific intensity on the sky, $\nu$ is the observed frequency, $\mathbfit{b}(t)$ is the baseline between the antennas used to form the visibility, $c$ is the speed of light in vacuum, $\hat{\mathbfit{n}}$ is a unit vector pointing towards different locations $(\theta,\phi)$ on the celestial sphere (with $\theta$ the polar angle and $\phi$ the azimuthal angle), $d\Omega$ is a differential solid angle element, and the integral is taken over the entire celestial sphere.
Note that the above equation implicitly bundles horizon effects into the beam, and the time variability is encoded in the baseline orientation and the beam's phase center.

We may expand the sky intensity in~\autoref{eq:measurement-equation} in spherical harmonics via
\begin{equation}
    \label{eq:sky-spherical-harmonics}
    I(\hat{\mathbfit{n}}) = \sum_{\ell,m} a_{\ell m} Y_\ell^m(\hat{\mathbfit{n}}),
\end{equation}
where $a_{\ell m}$ are the spherical harmonic coefficients and $Y_\ell^m(\hat{\mathbfit{n}})$, are the spherical harmonic functions normalized so that
\begin{equation}
    \label{eq:spherical-harmonic-definition}
    \int_{4\pi} Y_\ell^m(\hat{\mathbfit{n}}) Y_{\ell'}^{m'}(\hat{\mathbfit{n}})^* d\Omega = \delta_{\ell \ell'} \delta_{mm'}.
\end{equation}
We may additionally expand the beam-weighted fringe in spherical harmonics via
\begin{equation}
    \label{eq:beam-fringe-harmonic-expansion}
    A(\hat{\mathbfit{n}}, t) e^{-i2\pi\nu\mathbfit{b}(t)\cdot\hat{\mathbfit{n}}/c} = \sum_{\ell, m} K_{\ell m}(t) Y_\ell^m(\hat{\mathbfit{n}}),
\end{equation}
so that the beam transfer matrix $K_{\ell m}(t)$ is computed as
\begin{equation}
    \label{eq:beam-transfer-matrix}
    K_{\ell m}(t) = \int_{4\pi} A(\hat{\mathbfit{n}},t) e^{-i2\pi\nu\mathbfit{b}(t) \cdot \hat{\mathbfit{n}}/c} Y_\ell^m(\hat{\mathbfit{n}})^* d\Omega.
\end{equation}
Since we are assuming that the time evolution is strictly sourced by Earth rotation, evolving the visibilities in time amounts to applying an $m$-dependent phase so that
\begin{equation}
    \label{eq:time-dependent-beam-transfer-matrix}
    K_{\ell m}(t) = K_{\ell m}(t=0) e^{-im\omega_\oplus t},
\end{equation}
where $\omega_\oplus$ is the angular frequency of Earth's rotation.
For brevity, we will write the beam transfer matrix at $t = 0$ as $K_{\ell m}$.
Note that our definition of the beam transfer matrix differs from the definition in~\citet{Shaw:2014} by a normalization factor (\citealt{Shaw:2014} works with the area normalized beam, while we work with the peak normalized beam) and complex conjugation (\citealt{Shaw:2014} uses the conjugate expansion of the beam-weighted fringe).
Using both spherical harmonic expansions and the normalization condition allows us to rewrite the visibilities as
\begin{equation}
    \label{eq:visibility-from-harmonics}
    V(t) = \sum_{\ell, m} K_{\ell m} a_{\ell m}^* e^{-im\omega_\oplus t}.
\end{equation}
We may also opt to compute the sum over $\ell$ to express the visibility in terms of the $m$-modes $V_m$~\citep{Shaw:2014} as
\begin{equation}
    \label{eq:visibility-from-m-modes}
    V(t) = \sum_m V_m e^{-im\omega_\oplus t}.
\end{equation}

We now treat the cosmological 21-cm signal as a mean-zero Gaussian random field.
Under this assumption, the expected value of each $m$-mode vanishes, so we only need to compute $C(t,t') = \bigl\langle V(t) V(t')^* \bigr\rangle$.
Inserting~\autoref{eq:visibility-from-harmonics} into this expression yields
\begin{equation}
    C(t,t') = \sum_{\ell,m} \sum_{\ell',m'} K_{\ell m} K_{\ell'm'}^* \langle a_{\ell m}^* a_{\ell'm'} \rangle e^{-i\omega_\oplus (mt - m't')},
\end{equation}
which can be simplified as
\begin{equation}
    \label{eq:general-time-time-covariance}
    C(t,t') = \sum_{\ell,m} C_\ell \big|K_{\ell m}\big|^2 e^{-im\omega_\oplus (t-t')}, 
\end{equation}
where $C_\ell = \bigl\langle a_{\ell m} a_{\ell m}^* \bigr\rangle$ is the angular power spectrum of the cosmological signal.
Since the angular power spectrum of the cosmological signal evolves slowly with $\ell$~\citep{Santos:2006} relative to the sharply peaked beam transfer matrix, we may approximate $C_\ell \approx \sigma^2$ as constant and factor the angular power spectrum out of the sum, which leaves us with
\begin{equation}
    C(t,t') = \sigma^2 \sum_{\ell,m} \big|K_{\ell m}\big|^2 e^{-im\omega_\oplus(t-t')}.
\end{equation}
We now define the instrumental $m$-mode power spectrum $M_m$ as
\begin{equation}
    \label{eq:m-mode-power-spectrum}
    M_m \equiv \sum_\ell \big|K_{\ell m}\big|^2,
\end{equation}
which allows the time-time covariance to take on the simple form
\begin{equation}
    \label{eq:time-time-covariance}
    C(t,t') = \sigma^2 \sum_m M_m e^{-im\omega_\oplus(t-t')}.
\end{equation}
The time-time covariance in the observed cosmological signal is therefore uniquely characterized by the instrumental $m$-mode power spectrum.
For a more general case where $C_\ell$ cannot be treated as constant, however, one must instead compute the observed $m$-mode power spectrum $M_m = \sum_\ell C_\ell |K_{\ell m}|^2$.
Computing the $m$-mode power spectrum, instrumental or observed, is therefore a crucial step in obtaining the signal loss induced by a lossy time-based operation.

\subsubsection{The $m$-mode Power Spectrum}
\label{sec:M-MODE-SPECTRUM}
The $m$-mode power spectrum characterizes the time variability in the visibilities induced by spatially uncorrelated emission drifting through both the fringe pattern and the primary beam due to Earth rotation.
In a sense, the instrumental $m$-mode power spectrum describes the intrinsic time variability of the instrument, since it characterizes how the data is correlated in time in the absence of correlated structures on the sky---in other words, the $m$-mode power spectrum characterizes the correlation structure of the data across time when the time evolution is completely determined by the instrument response.
This has two useful consequences: First, the $m$-mode power spectrum can be used to design filters tuned to reject signals not consistent with the instrument response; Second, the $m$-mode power spectrum can be used as a starting point for calculating signal loss from a time-based filter, either as an approximation for sky-based signals with nontrivial spatial correlations or as an analytic expectation for spatially uncorrelated signals.
We explore both of these consequences (although omitting an analysis in the context of correlated structures on the sky) further in Section~\ref{sec:SIGNAL-LOSS}.

\begin{figure}        \includegraphics[width=\columnwidth]{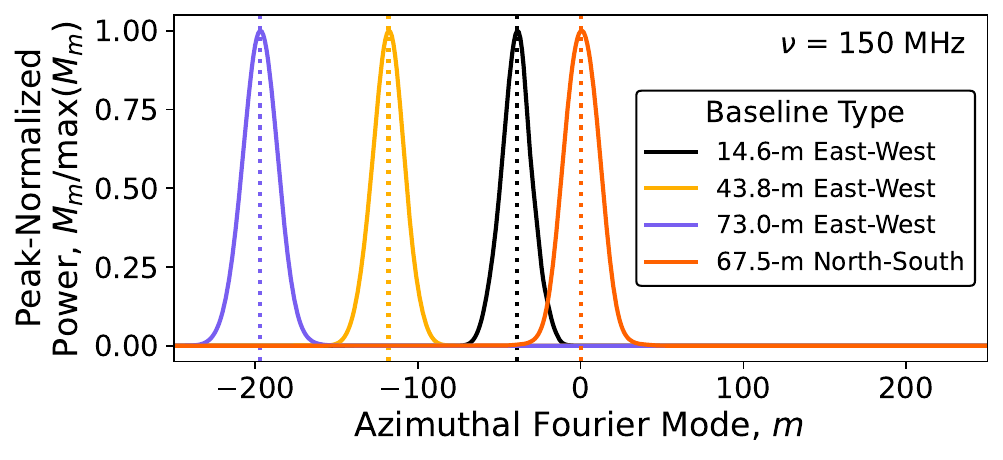}
    \caption{
        Peak-normalized $m$-mode power spectra for three East-West baselines and one North-South baseline.
        The vertical dotted lines indicate the highest sensitivity $m$-mode predicted by the ``instantaneous fringe-rate''~\citep{Parsons:2016} at the most sensitive part of the primary beam.
        Baselines with longer East-West projected lengths are sensitive to fluctuations on finer angular scales in the East-West direction, and hence their $m$-mode power spectra peak at higher azimuthal Fourier modes $|m|$.
    }
    \label{fig:m-mode-spectra}
\end{figure}

In~\autoref{fig:m-mode-spectra} we show $m$-mode power spectra for four representative baselines at 150 MHz (three baselines are oriented East-West and one is oriented North-South).
The $m$-mode power spectra were computed using numerical electromagnetic simulations of the HERA beam~\citep{Fagnoni:2021} for the Phase II instrument~\citep{Berkhout:2024}, with $M_m$ and $K_{\ell m}$ computed according to~\autoref{eq:m-mode-power-spectrum} and~\autoref{eq:beam-transfer-matrix}, respectively.
The generic trend seen in~\autoref{fig:m-mode-spectra} is that the $m$-mode power spectrum peaks at larger $|m|$ as the projected East-West baseline length increases, and this trend agrees with intuition from the ``instantaneous fringe-rate'' picture proposed by~\citet{Parsons:2016}.
Since the fringe pattern oscillates in the same direction as the baseline orientation, and since the fringe period decreases with increasing baseline length, visibilities measured by drift scan telescopes will fluctuate more rapidly in time for baselines with longer East-West projected lengths.
Baselines with longer East-West projected lengths are therefore more sensitive to fluctuations on shorter timescales and thus higher $|m|$.

\subsection{Covariance in the Fringe-Rate Domain}
\label{sec:FRINGE-RATE-COVARIANCE}

While working in the $m$-mode basis is extremely helpful for making problems analytically tractable, it is often not possible to access this basis from realistic observational data.
Since the $m$-mode expansion uses harmonics of the Earth's rotation as a basis, we can only recover the $m$-modes through a Fourier transform for observations that cover a full sidereal day.
For observations that cover less than a sidereal day, the Fourier transformed data is instead a function of \emph{fringe-rate}, which have a nontrivial correlation structure (relative to the diagonal $m$-mode covariance) that we explore in the following paragraphs.

We define the fringe-rate transform of the visibilities $\bar{V}(f_r)$ as a tapered Fourier transform, so that
\begin{equation}
    \bar{V}(f_r) = \int w(t) V(t) e^{-i2\pi f_r t} dt,
    \label{eq:fringe-rate-transform}
\end{equation}
where $w(t)$ encodes either the integral cutoff for a non-tapered transform over a non-periodic observation or the apodization used to suppress sidelobes that come from a non-periodic boundary.
The covariance in the fringe-rate domain can be computed in the same manner as in the time domain, via
\begin{equation}
    C(f_r, f_r') = \bigl\langle \bar{V}(f_r) \bar{V}(f_r')^* \bigr\rangle.
\end{equation}
We can relate the covariance in the fringe-rate domain to the $m$-mode power spectrum by rewriting the visibilities in the fringe-rate domain in terms of $m$-modes.
Inserting the $m$-mode expansion (\autoref{eq:visibility-from-m-modes}) into~\autoref{eq:fringe-rate-transform} yields
\begin{equation}
    \bar{V}(f_r) = \sum_m V_m \bar{w}\biggl( f_r + \frac{m\omega_\oplus}{2\pi}\biggr),
    \label{eq:fringe-rates-and-m-modes}
\end{equation}
where $\bar{w}(f_r)$ is the Fourier transform of the taper $w(t)$.
Note that this mathematically encodes the fact that the $m$-modes can only be obtained through a Fourier transform in the limit of a full sidereal day of observation.
Performing a Fourier transform over a subset of a day spanning time $t_1$ to time $t_2$ without a taper is equivalent to using a top-hat function $\mathcal{T}(t)$ for the taper, where
\begin{equation}
    \label{eq:top-hat-defn}
    \mathcal{T}(t) = \begin{cases} 1, \quad t_1 \leq t \leq t_2 \\ 0, \quad {\rm else}
    \end{cases}.
\end{equation}
In this limit,
\begin{equation}
    \label{eq:sinc-sidelobes}
    \bar{w}(f_r) = (t_2 - t_1) e^{-i\pi f_r(t_1+t_2)} {\rm sinc}\bigl(\pi f_r (t_2 - t_1)\bigr),
\end{equation}
where ${\rm sinc}(x) \equiv \sin(x)/x$, and so the fringe-rate modes obtained through the Fourier transform are nontrivial linear combinations of the $m$-modes (i.e., the fringe-rate modes are obtained by convolving the $m$-modes with a sinc convolution kernel whose width depends on the length of the observation $|t_2 - t_1|$).
Using~\autoref{eq:fringe-rates-and-m-modes}, we can compute the covariance between different fringe-rate modes via
\begin{equation}
    C\bigl(f_r, f_r'\bigr) = \sum_m \tilde{w}\biggl( f_r + \frac{m\omega_\oplus}{2\pi}\biggr) M_m \tilde{w} \biggl( f_r' + \frac{m\omega_\oplus}{2\pi}\biggr)^*.
    \label{eq:fringe-rate-covariance}
\end{equation}
The fringe-rate covariance is therefore obtained through a linear transformation of the $m$-mode covariance.

\begin{figure}
\includegraphics[width=\columnwidth]{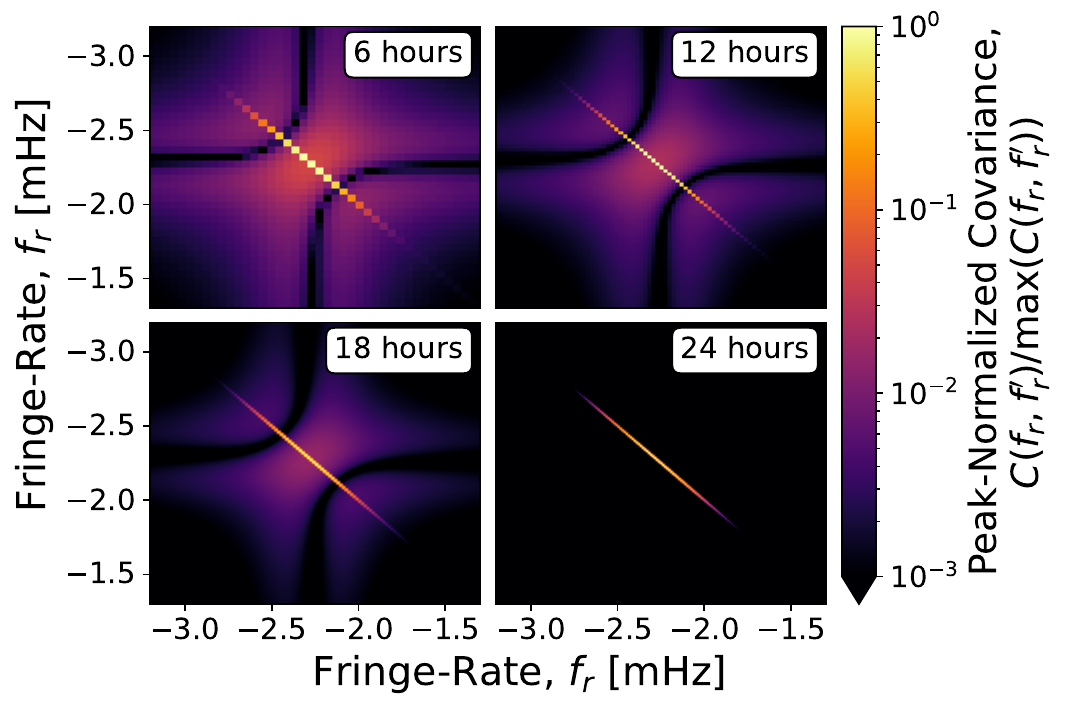}
    \caption{
        Fringe-rate covariance a 73-meter East-West baseline at 150 MHz for a few different observing periods, with the observation duration noted in the text box at the upper right of each panel.
        Each panel only shows a subset of the fringe-rate covariance, centered on the peak of the fringe-rate profile---note that these are not centered at zero fringe-rate.
        Shorter observing periods result in stronger correlations between neighboring fringe-rate modes, as well as longer correlation lengths.
        In the limit of a full day of observation, however, the fringe-rate covariance is diagonal.
    }
    \label{fig:fringe-rate-cov}
\end{figure}

Example covariance matrices in the fringe-rate domain are shown for various lengths of observation time with uniform weighting (i.e., $w(t) = \mathcal{T}(t)$ with varying values of $|t_2 - t_1|$ in~\autoref{eq:top-hat-defn}) in~\autoref{fig:fringe-rate-cov}.
\autoref{fig:fringe-rate-cov} shows that the fringe-rate covariance is mostly diagonal across a broad range of total observation times $|t_2 - t_1|$ and that most of the power is concentrated around fringe-rates near $f_r = m_0\omega_\oplus/2\pi$, where $m_0$ is the $m$-mode at which the $m$-mode power spectrum peaks.
The off-diagonal elements tend to decrease in amplitude with increasing observation duration and ultimately vanish in the limit of a full sidereal day of observation.\footnote{The hyperbolic dark contours in~\autoref{fig:fringe-rate-cov} are a characteristic feature associated with taking the Fourier transform of a diagonal matrix that is not proportional to the identity. In this context, we are effectively taking the Fourier transform of the instrumental $m$-mode power spectrum to obtain the time-time covariance.}
This may be thought of as a transition from an ``incomplete'' basis to a complete basis for describing temporal structure over a full sidereal day in the following sense: the temporal structure in drift-scanning data is sourced from Earth rotation, and therefore the fundamental mode is tied to Earth's rotational period; for observations that span less than a sidereal day, the fringe-rate transform of the time series data is effectively a harmonic expansion using the wrong fundamental mode, thereby inducing correlations between different fringe-rate modes.
The $\tilde{w}(f_r + m\omega_\oplus/2\pi)$ term in~\autoref{eq:fringe-rates-and-m-modes} and~\autoref{eq:fringe-rate-covariance} may therefore be interpreted as a mode-mixing matrix that encodes the covariance between different fringe-rate modes originating from incomplete coverage of a full sidereal day.
\autoref{eq:fringe-rate-covariance} thus provides us with a method of decoupling the instrument properties from the details of a particular observing campaign when computing the fringe-rate covariance, since the shape of the $m$-mode power spectrum $M_m$ only depends on instrument properties and the mode-mixing matrix $\tilde{w}(f_r + m\omega_\oplus/2\pi)$ depends only on the choice of apodization and the observation duration.

\subsubsection{Fringe-Rate Profiles}
\label{sec:FRATE-PROFILES}

\begin{figure}
\includegraphics[width=\columnwidth]{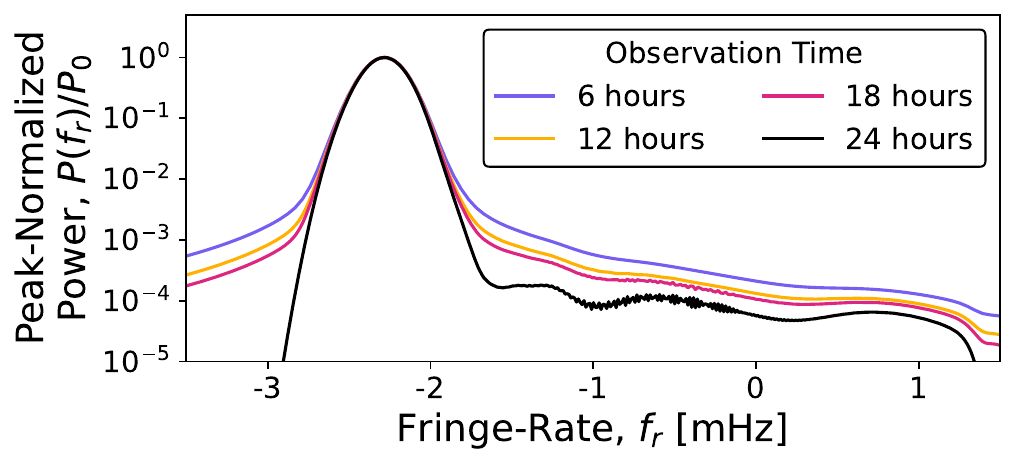}
    \caption{
        Peak-normalized fringe-rate profiles for a 73-meter East-West baseline, estimated at 150 MHz for various lengths of total observing time.
        As the total amount of observing time approaches a sidereal day, the sinc sidelobes associated with a non-periodic observing window become increasingly suppressed, and entirely vanish when a full sidereal day is used to compute the fringe-rate profile.
    }
    \label{fig:fringe-rate-profiles}
\end{figure}

The diagonal of the fringe-rate covariance, which we refer to as the \emph{fringe-rate profile}, is shown in~\autoref{fig:fringe-rate-profiles} for the four covariance matrices from~\autoref{fig:fringe-rate-cov}.
The generic trend seen in~\autoref{fig:fringe-rate-profiles} is that the peak of the fringe-rate profile is independent of the observation duration, and shorter total observation times produce larger ``wings'' in the fringe-rate profiles.
The enhanced wings in the fringe-rate profiles is in alignment with intuition from Fourier analysis in the sense that the fringe-rate profile $P(f_r)$ is related to the $m$-mode power spectrum through a convolution,
\begin{equation}
    \label{eq:fringe-rate-profile}
    P(f_r) = C(f_r, f_r) = \sum_m \Bigg| \tilde{w}\biggl(f_r + \frac{m\omega_\oplus}{2\pi}\biggr) \Bigg|^2 M_m. 
\end{equation}
Shorter total observation times lead to wider sinc convolution kernels, which result in broader wings in the fringe-rate profiles.
In addition to this, short observation times can lead to subtle discretization issues, as suggested by the fact that most of the power in the fringe-rate profile is contained in roughly ten fringe-rate bins for the six hour observation, as seen in the upper-left panel of~\autoref{fig:fringe-rate-cov}.
The effects of a short observation duration are thus twofold: on the one hand, the non-periodic boundary scatters power away from the peak, which causes a larger fraction of the total power to be contained in the wings of the fringe-rate profile; on the other hand, a shorter observation duration results in coarser fringe-rate resolution, which can lead to a majority of the power being contained in just a few fringe-rate bins near the peak of the fringe-rate profile.
These effects are subtle, but may be important to consider in detailed analyses.

\begin{figure}
    \includegraphics[width=\columnwidth]{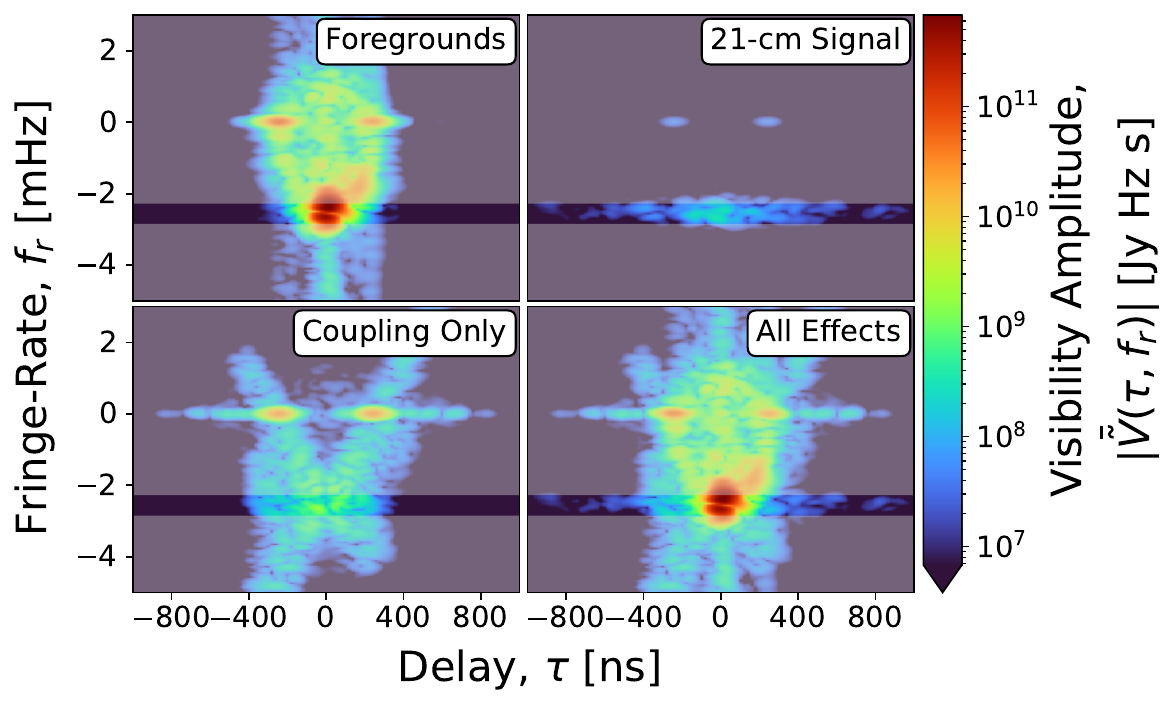}
    \caption{
        Comparison of how mutual coupling, foregrounds, and a mock 21-cm signal manifest in the visibilities for a baseline with a long East-West projection, shown in fringe-rate versus delay space.
        The greyed out region indicates modes that would be excised by a fringe-rate filter designed to mitigate systematics while retaining a majority of the cosmological 21-cm signal as described in Section~\ref{sec:FRF-IMPLEMENTATION}.
        Note that the bulk of the 21-cm signal is confined to a narrow range of fringe-rates, while the foregrounds and mutual coupling span a much broader range of fringe-rates.
    }
    \label{fig:vis-breakdown}
\end{figure}

Since the cosmological 21-cm signal is statistically isotropic and fairly uncorrelated on degree scales, the $m$-mode power spectra---and hence the fringe-rate profiles---may be used to design fringe-rate filters that retain a majority of the cosmological signal while rejecting other undesired features in the data, such as mutual coupling~\citep{Kern:2019,Kern:2020a,Josaitis:2022,Rath&Pascua:2024} and the ``pitchfork'' effect from bright diffuse emission on the horizon~\citep[e.g., ][]{Thyagarajan:2015,Charles:2023}.
In~\autoref{fig:vis-breakdown}, we show simulated visibilities from~\citet{Rath&Pascua:2024} in the fringe-rate versus delay domain, $\tilde{\bar{V}}(\tau,f_r)$, which are obtained from the visibility ``waterfalls'' $V(\nu,t)$ by performing a tapered Fourier transform along the time axis (as in~\autoref{eq:fringe-rate-transform}) as well as a tapered Fourier transform along the frequency axis via
\begin{equation}
    \label{eq:delay-transform}
    \tilde{V}(\tau) = \int B(\nu) V(\nu) e^{-i2\pi\nu\tau} d\nu,
\end{equation}
where $B(\nu)$ is the frequency taper and the delay $\tau$ is the Fourier dual to frequency $\nu$.
\autoref{fig:vis-breakdown} shows how the cosmological 21-cm signal can be discriminated against mutual coupling and the pitchfork based on the fringe-rate modes that the signals occupy:
The upper-right panel of~\autoref{fig:vis-breakdown} shows that the majority of the cosmological signal is confined to a narrow range of fringe-rates, in accordance with the expected fringe-rate profile computed from the $m$-mode power spectrum;
The upper-left panel shows that the foreground signal occupies a wider range of fringe-rates, with two bright regions centered around zero fringe-rate at the horizon delay (i.e., the delay corresponding to the light travel time between the two antennas used to form the visibility) which form the pitchfork;
The bottom-left panel shows that mutual coupling extends over an even broader range of fringe-rates and delays, with the bulk of the coupling signal (for the particular range of times used in the simulation) coming from around zero fringe-rate.
By removing the greyed out regions of~\autoref{fig:vis-breakdown}, it is thus possible to reduce contaminants from coupling effects and widefield foreground effects at delays that would otherwise be dominated by the cosmological 21-cm signal.
We describe how such a filter may be constructed from the fringe-rate profiles in Section~\ref{sec:FRF-IMPLEMENTATION}.

\subsection{Comparison to Previous Work}
\label{sec:PREV-WORK}
\citet{Parsons:2016} previously explored the effects of fringe-rate filters on interferometric data, framing the problem as one of ``beam sculpting'' in the sense that fringe-rate filters can be interpreted as operations that modify the shape of the primary beam.
This interpretation is based on the association between locations on the sky $\hat{\bf n}$ and \emph{instantaneous fringe-rates} $f_r(\bvec{n})$ via
\begin{equation}
    f_r(\bvec{n}) = \frac{\nu (\boldsymbol{\omega}_\oplus \times \mathbfit{b}) \cdot \bvec{n}}{c},
    \label{eq:instantaneous-fringe-rates}
\end{equation}
where $\boldsymbol{\omega}_\oplus$ is the Earth's rotational angular velocity vector (i.e., the vector form of $\omega_\oplus$ introduced in~\autoref{eq:time-dependent-beam-transfer-matrix}).
This association between fringe-rates and locations on the sky comes with a natural interpretation of fringe-rate filters modifying the primary beam, since a fringe-rate filter effectively suppresses the instrument's sensitivity to different locations on the sky.
A convenient application of this interpretation is that any signal loss induced by a fringe-rate filter can be captured through an effective primary beam, and power spectra formed with fringe-rate filtered visibilities may be normalized by using this effective primary beam to compute the beam integrals.

While the beam sculpting interpretation is conceptually straightforward, it is insufficient for experiments like HERA where the short baselines used for cosmology have lengths comparable to the dish diameter.
The instantaneous fringe-rate model was derived in~\citet{Parsons:2016} by assuming that the time variability in the data was dominated by the sky drifting through the interferometric fringe and neglecting any variability associated with the sky drifting through the primary beam.
In other words, locations on the sky can only be mapped to particular fringe-rates when the fringe period (the time it takes a source to drift through one period of the fringe pattern) is much shorter than the beam crossing time (how long it takes a source to drift through the main lobe of the primary beam)---when the fringe period is comparable to the beam crossing time, locations on the sky can no longer be mapped to individual fringe-rates.
This constraint is equivalent to requiring that the baseline under consideration is substantially longer than the dish diameter, as can be shown by appealing to the case of an Airy beam.
For an Airy beam, the main lobe width scales like $\lambda/D$, where $D$ is the dish diameter and $\lambda = c/\nu$ is the observed wavelength, while the angular period of the interferometric fringe near zenith for an East-West oriented baseline generically scales like $\lambda / b$, where $b$ is the baseline length.
For drift scan telescopes, sources move through the fringe and the beam at the same rate, so the ratio of the fringe period to the beam crossing time scales like $D/b$ and hence the fringe period is comparable to the beam crossing time for baselines that have projected East-West lengths comparable to the dish diameter.
For arrays like HERA, whose sensitivity is dominated by short baselines that are at most a few dish diameters in length, the instantaneous fringe-rate model is therefore insufficient for characterizing the instrument's fringe-rate response, and so the filter-induced signal loss cannot generically be captured by a simple modification to the beam integrals.

\begin{figure*}
    \includegraphics[width=\textwidth]{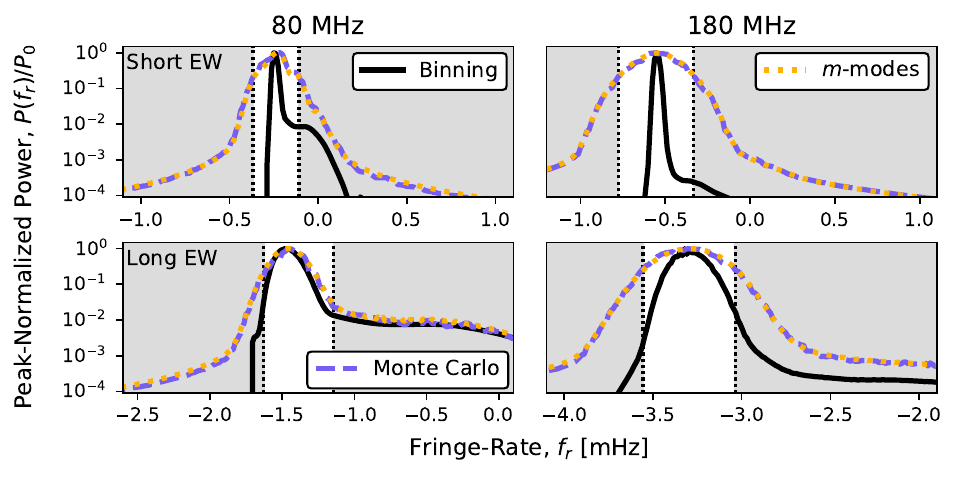}
    \caption{
        Comparison between various ways of estimating fringe-rate profiles for two representative baselines from HERA.
        The top row shows fringe-rate profiles for a short (15-meter) East-West baseline, while the bottom row shows profiles for a long (88-meter) East-West baseline.
        The left column shows the profiles at 80 MHz, while the right column shows the profiles at 180 MHz.
        The solid black line estimates the fringe-rate profiles using the histograming procedure described in Section~\ref{sec:PREV-WORK}.
        The dashed purple line estimates the fringe-rate profiles from the visibility simulations used in the Monte Carlo analysis described in Section~\ref{sec:SIMULATIONS}.
        The dotted yellow line is obtained by taking the diagonal of the fringe-rate covariance matrix computed in Section~\ref{sec:FRINGE-RATE-COVARIANCE}.
        The vertical dotted lines show the boundaries of a fringe-rate filter designed to retain 90\% of the cosmological signal, computed by integrating the dotted yellow fringe-rate profiles; the shaded grey region shows the associated filter rejection regions.
        There are two salient features of the fringe-rate profiles shown in this figure: First, the fringe-rate profiles computed from our $m$-mode-based analysis are in excellent agreement with the fringe-rate profiles numerically estimated from the suite of visibility simulations; Second, the histogramming method for estimating the fringe-rate profiles tends to predict much narrower fringe-rate profiles than either of the two other methods predict.
        This suggests that our $m$-mode-based formalism should accurately predict the signal loss, and that the signal loss estimated using the histogramming approach would be substantially lower than the actual signal loss. 
    }
    \label{fig:frate-profile-comparisons}
\end{figure*}

\autoref{fig:frate-profile-comparisons} compares three sets of fringe-rate profiles for two different types of baselines at two different frequencies for the Phase II HERA instrument.
One set of fringe-rate profiles is computed using a method based on \citet{Parsons:2016}, whereby locations on the sky are assigned instantaneous fringe-rates, weighted by the square of the primary beam, and then binned as a function of fringe-rate following the prescription from HERA Memo 111~\citep{Ewall-Wice:2022};
Another set of fringe-rate profiles is computed using the $m$-mode formalism according to the prescription in Section~\ref{sec:FRATE-PROFILES};
The third set of fringe-rate profiles are estimated from the Monte Carlo simulations used in Section~\ref{sec:SIMULATIONS}.
Both of the baselines we consider are purely East-West oriented, with one baseline having a length similar to the diameter of one HERA dish and the other being many ($\sim$5) times longer.
We estimate the fringe-rate profiles at 80 MHz and 180 MHz to show how the quality of the estimates made using the formalism based on \citet{Parsons:2016} depends on the spectral evolution of the primary beam width.

\autoref{fig:frate-profile-comparisons} clearly shows that the fringe-rate profiles estimated from the instantaneous fringe-rate model do not correctly characterize the fringe-rate response of the instrument when the baseline length is comparable to the dish diameter, while the profiles computed with the $m$-mode formalism show good agreement with numerical simulation in all cases.
In particular, the fringe-rate profile estimates based on the instantaneous fringe-rate model (which we will call the ``instantaneous fringe-rate profiles'') are universally narrower than the true fringe-rate profiles, which has two crucially important consequences.
First, fringe-rate filters designed by appealing only to the instantaneous fringe-rate profiles will attenuate more of the cosmological signal than intended.
Second, signal loss calculations performed using the instantaneous fringe-rate profiles will under-predict the actual amount of signal loss incurred by a fringe-rate filter.
While the instantaneous fringe-rate profile tends to be more narrow than the true fringe-rate profile, the exact level of inaccuracy in the instantaneous fringe-rate profile depends on beam chromaticity and baseline, as demonstrated by the bottom row of~\autoref{fig:frate-profile-comparisons}.
For HERA antennas at 80 MHz, the instantaneous fringe-rate profile for a long East-West baseline only slightly disagrees with the true fringe-rate profile, while the instantaneous fringe-rate profile for same baseline at 180 MHz is noticeably narrower than the true fringe-rate profile.
Evidently, the HERA beam narrows more rapidly as a function of frequency than the fringe spacing narrows, and so the instantaneous fringe-rate approximation breaks down at high frequencies even though it holds at low frequencies.
The main takeaway is that while the approximate method based on~\citet{Parsons:2016} may be appropriate for characterizing the fringe-rate response of an array of widefield antennas without short baselines (such as PAPER), the instantaneous fringe-rate profile is generally a poor representation of the actual fringe-rate response of the instrument; however, the instantaneous fringe-rate approximation does yield an accurate prediction of the location of the peak in the fringe-rate profile.
We therefore recommend using the exact $m$-mode calculations to obtain the fringe-rate response of a drift-scanning interferometer, but note that applications solely concerned with finding the peak of the fringe-rate response may instead use the instantaneous fringe-rate formalism.

\section{Signal Loss}
\label{sec:SIGNAL-LOSS}
In Section~\ref{sec:TIME-TIME-COVARIANCE} we reviewed how the time variability in interferometric data is uniquely characterized by the $m$-mode power spectrum via~\autoref{eq:time-time-covariance}.
In Section~\ref{sec:PREV-WORK}, we showed how the instantaneous fringe-rate model is insufficient for characterizing the time variability in the data, in particular for baselines with lengths comparable to the dish diameter.
In this section, we take the lessons learned from Section~\ref{sec:PREV-WORK} and the tools reviewed in Section~\ref{sec:TIME-TIME-COVARIANCE} to extend the work of~\citet{Parsons:2016} and provide a more general technique for computing the signal loss associated with a lossy time-based operation.

\subsection{Linear Filters, Transfer Functions, and Expected Loss}
\label{sec:SIGNAL-LOSS-FORMALISM}
For this work, we will limit our scope to linear operators $\mathbf{T}$ (which we will interchangeably refer to as ``filters'') acting on time-ordered visibilities $\mathbfit{V}$ to produce ``filtered'' visibilities $\mathbfit{V}'$ via
\begin{equation}
    \label{eq:general-linear-operator}
    \mathbfit{V}' = \mathbf{T} \mathbfit{V},
\end{equation}
since the most commonly employed time-based operations (e.g., fringe-rate filtering or coherent time averaging) tend to be linear operations.
Determining the expected signal loss associated with the operator $\mathbf{T}$ amounts to computing the ratio of the expected power after applying the operation to the expected power before applying the operation.
In this paper, we will restrict our attention to a power spectrum estimator that employs an incoherent weighted average via
\begin{equation}
    \label{eq:power-spectrum-estimate}
    \hat{P} = \mathcal{Q} \sum_t w_t |V_t|^2,
\end{equation}
where $\hat{P}$ is the estimated average power, $\mathcal{Q}$ is a normalization factor that handles the conversion from telescope units to cosmological units, $V_t$ is the visibility at time $t$, and $w_t$ is the weight with $\sum_t w_t = 1$.
Note that the average power in~\autoref{eq:power-spectrum-estimate} is computed frequency-by-frequency, rather than through a delay spectrum estimate as in Section~\ref{sec:MC-SIGNAL-LOSS}.
In Section~\ref{sec:MC-SIGNAL-LOSS}, we provide a method for adapting the tools from this section to a delay spectrum estimator.
Assuming the noise variance is uniform across time and a uniform number of samples per integration, the average power is an incoherent average computed as
\begin{equation}
    \label{eq:inverse-variance-weighted-avg}
    \hat{P} = \frac{\mathcal{Q}}{N_t} \sum_t |V_t|^2,
\end{equation}
where $N_t$ is the number of integrations incoherently averaged together.
The expected average power $P \equiv \langle \hat{P} \rangle$ is just the ensemble average of~\autoref{eq:inverse-variance-weighted-avg}, which can be expressed in terms of the time-time covariance $\mathbf{C}$ as
\begin{equation}
    \label{eq:average-power}
    P = \frac{\mathcal{Q}}{N_t} {\rm tr}\mathbf{C},
\end{equation}
where ${\rm tr}(\cdots)$ is the trace operator, since $\sum_t |V_t|^2 = {\rm tr}(\mathbfit{VV}^\dagger)$ and $\langle \mathbfit{VV}^\dagger \rangle = \mathbf{C}$.
For the remainder of this paper, we will limit our scope to uniformly weighted power spectrum estimates (i.e.,~\autoref{eq:inverse-variance-weighted-avg} and~\autoref{eq:average-power}).

We may compute the expected average power $P'$ in the filtered visibilities by replacing the unfiltered visibilities in~\autoref{eq:inverse-variance-weighted-avg} with the filtered visibilities in~\autoref{eq:general-linear-operator}.
Performing this substitution and taking the expectation value, we obtain
\begin{equation}
    \label{eq:average-filtered-power}
    P' = \frac{\mathcal{Q}}{N'_t} {\rm tr}\Bigl( \mathbf{TCT}^\dagger \Bigr),
\end{equation}
where $N'_t$ is the number of times in the filtered visibilities, which may be different from $N_t$ depending on the type of operation applied to the data.
The fraction of power retained after applying the filter is just the ratio of~\autoref{eq:average-filtered-power} and~\autoref{eq:average-power}, so the expected signal loss $L$ incurred by the filter $\mathbf{T}$ is
\begin{equation}
    \label{eq:expected-signal-loss}
    L = 1 - \frac{N_t}{N'_t} \frac{{\rm tr}\Bigl( \mathbf{TCT}^\dagger \Bigr)}{{\rm tr}\mathbf{C}}.
\end{equation}
Using the cyclic property of the trace to permute the terms in the numerator, the loss may also be written as
\begin{equation}
    \label{eq:signal-loss-cycled}
    L = 1 - \frac{N_t}{N'_t} \frac{{\rm tr}\Bigl(\mathbf{T}^\dagger \mathbf{TC}\Bigr)}{{\rm tr}\mathbf{C}}.
\end{equation}
If the operator $\mathbf{T}$ is unitary, then $\mathbf{TT}^\dagger = \mathbf{I}$, $N'_t = N_t$, and the operation is lossless, as expected.

The signal loss calculation provided in~\autoref{eq:expected-signal-loss} can be performed in other bases that are related to the time domain through an \emph{invertible} linear transformation $\mathbf{R}$, whereby the visibilities in the new basis $\mathbf{V}_\mathcal{R}$ are related to the time-ordered visibilities $\mathbf{V}$ via
\begin{equation}
    \label{eq:rotated-visibilities}
    \mathbf{V}_\mathcal{R} = \mathbf{RV}.
\end{equation}
Applying the same transformation to~\autoref{eq:general-linear-operator} and inserting $\mathbf{R}^{-1} \mathbf{R}$ between the filter and the time-ordered visibilities, the filtered visibilities in the new basis $\mathbf{V}'_\mathcal{R}$ are computed via
\begin{equation}
    \mathbf{V}'_\mathcal{R} = \mathbf{RTR}^{-1} \mathbf{RV} = \mathbf{T}_\mathcal{R} \mathbf{V}_\mathcal{R},
\end{equation}
where $\mathbf{T}_\mathcal{R} = \mathbf{RTR}^{-1}$ is the filter in the new basis.
Propagating this through to the signal loss calculation, we obtain
\begin{equation}
    \label{eq:signal-loss-general-basis}
    L = 1 - \frac{N_t}{N'_t} \frac{{\rm tr}\Bigl( \mathbf{T}_\mathcal{R} \mathbf{C}_\mathcal{R} \mathbf{T}_\mathcal{R}^\dagger \Bigr)}{{\rm tr} \mathbf{C}_\mathcal{R}},
\end{equation}
where $\mathbf{C}_\mathcal{R} = \mathbf{RCR}^{-1}$ is the covariance in the new basis.
The signal loss calculation may therefore be computed in any basis desired, as long as the filter and covariance are appropriately transformed to that basis.
It is crucially important, however, that the time samples contain enough information to reconstruct the data in the basis used for computing the signal loss.
For example, if the signal loss calculation is performed in the $m$-mode basis when the data does not span a full sidereal day, then the mode-mixing effects of the incomplete sampling must be included in the calculation of $\mathbf{T}_\mathcal{R}$.

\autoref{eq:signal-loss-general-basis} has a particularly useful computational implication: if we choose a basis where the off-diagonal terms in either the filter or the covariance are small (i.e., $A_{ij} / \sqrt{|A_{ii} A_{jj}|} \ll 1$, where $\mathbf{A}$ is either $\mathbf{T}$ or $\mathbf{C}$), then we can greatly reduce the computational cost of calculating the signal loss.
Since the off-diagonal structure in the fringe-rate covariance tends to be small (see~\autoref{fig:fringe-rate-cov}), it is often helpful to perform the signal loss calculation in the fringe-rate domain, using only the fringe-rate profile (\autoref{eq:fringe-rate-profile}) and the ``filter transfer matrix'' $\bar{\mathbf{T}} = \mathbf{FTF}^{-1}$, where $\mathbf{F}$ is the Fourier transform operator.
When performing the calculation in the fringe-rate domain, however, it is crucially important to consider the structure of the filter transfer matrix---many filter implementations have significant off-diagonal structure in the fringe-rate domain, and in these cases one must use the entire filter transfer matrix in the signal loss calculation.

\begin{figure}
    \includegraphics[width=\columnwidth]{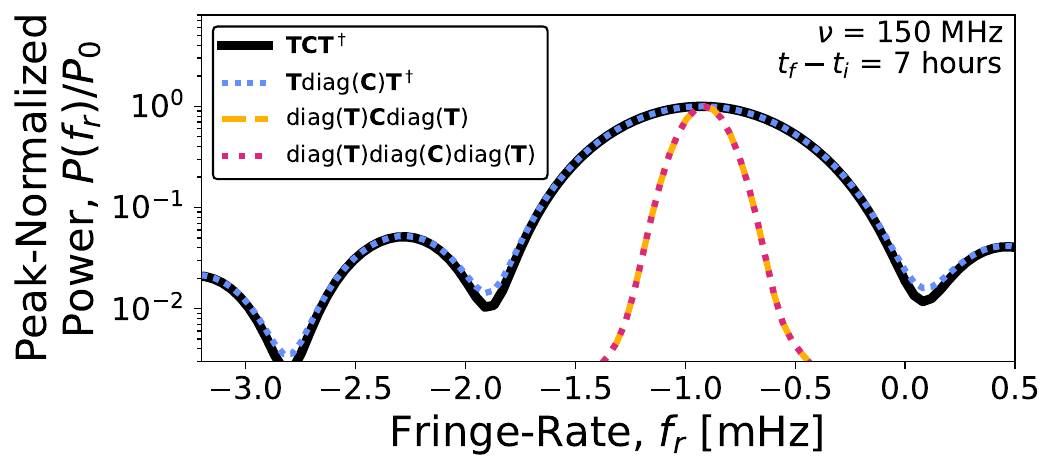}
    \caption{
        Comparison of filtered fringe-rate profiles computed four ways:
        In solid black, we show the computation using the full matrix product;
        In densely dotted blue, we show the computation using the full filter transfer matrix but only the fringe-rate profile;
        In dashed yellow, we show the computation using the full fringe-rate covariance but only the diagonal of the filter transfer matrix;
        In dotted pink, we show the computation using only the fringe-rate profile and the diagonal of the filter transfer matrix.
        The data shown here are for a 29-m East-West baseline at 150 MHz.
        The filter used here is the ``main lobe'' fringe-rate filter~\citep{Rath&Pascua:2024} designed to retain at least 90\% of the observed cosmological power.
        The two filtered fringe-rate profiles computed using the full filter transfer matrices (solid black and dotted blue) are in good agreement with each other and poor agreement with the filtered fringe-rate profiles computed using only the diagonal of the filter transfer matrix.
        This indicates that the off-diagonal entries in the fringe-rate covariance do not contribute much to the signal loss calculation, but the off-diagonal entries in the filter transfer matrix may strongly contribute to the signal loss calculation.
    }
    \label{fig:filtered-profile-comparison}
\end{figure}

In~\autoref{fig:filtered-profile-comparison}, we show how ignoring the off-diagonal structure in the fringe-rate covariance and filter transfer matrix manifests in the filtered fringe-rate profile.
We perform this comparison for the ``main lobe'' filter~\citep{Rath&Pascua:2024} that is used to mitigate mutual coupling effects in HERA data.
The filter used in generating~\autoref{fig:filtered-profile-comparison} is designed as a top-hat filter in fringe-rate, but the filter is implemented in a way that creates non-negligible off-diagonal structure in the filter transfer matrix (see the bottom right panel of~\autoref{fig:dpss-filter-matrices} for an example).
The off-diagonal structure in the filter transfer matrix clearly matters in this context, since the calculations that only involve the diagonal of the filter transfer matrix strongly disagree with those that use the full filter transfer matrix.
Conversely, the off-diagonal structure in the fringe-rate covariance does not matter much (as expected based on~\autoref{fig:fringe-rate-cov}), since the filtered fringe-rate profiles using the full filter transfer matrix are generally in good agreement with one another, with only small differences between the result using the fringe-rate profile and the result using the full fringe-rate covariance.

\subsection{Filter Examples}
With the signal loss formalism established, we now shift our attention to some specific examples of commonly used fringe-rate filters.
In this section, we will discuss coherent time averaging and filtering with Discrete Prolate Spheroidal Sequences~\citep[DPSS,][]{Slepian:1978,Percival&Walden:1993,Ewall-Wice:2020} as two example applications.
Notably, neither of these operations act on the visibilities in the fringe-rate domain; however, since they are both time-based operations, they may equivalently be thought of as fringe-rate filters.

\subsubsection{DPSS Filters}
\label{sec:DPSS-FILTERS}
DPSS-based filtering techniques have gained traction in recent years~\citep[e.g.,][]{Ewall-Wice:2020}, since the DPSS-filtered products are intrinsically smooth and are compact over a user-defined range of Fourier modes (i.e., DPSS-filtered products have very little leakage outside of some predetermined set of contiguous Fourier modes).
A DPSS-based filter is implemented as a \emph{least-squares filter}, whereby the filtered visibilities are obtained through a least-squares fit to an orthonormal set of vectors $\{\phi_k(t)\}$ called ``DPSS modes''.
The set of DPSS modes are uniquely determined by the filter half-width $\delta f_r$, the number of samples $N_t$ in the time series, and the integration time $\delta t$.
Given these parameters, the DPSS modes can be computed by solving the eigenvalue equation~\citep{Slepian:1978}
\begin{equation}
    \label{eq:dpss-eigenvalue-equation}
    \sum_{j=0}^{N_t-1} \frac{\sin\bigl(2\pi\delta f_r \delta t(i - j)\bigr)}{\pi(i - j)} \phi_k(t_j) = \lambda_k \phi_k(t_i),
\end{equation}
where $\lambda_k$ is the \emph{spectral concentration} and may equivalently be computed via
\begin{equation}
    \label{eq:spectral-concentration}
    \lambda_k \equiv \sum_{|f_r| \leq \delta f_r} |\bar{\phi}_k(f_r)|^2,
\end{equation}
where $\bar{\phi}_k(f_r)$ is the fringe-rate transform of the DPSS mode $\phi_k(t)$.
The DPSS modes are centered on zero fringe-rate and the spectral concentration $\lambda_k$ measures the fraction of power contained on the interval $f_r \in [-\delta f_r, \delta f_r]$; higher order DPSS modes have smaller spectral concentrations, so $\lambda_i < \lambda_j$ for $i > j$.

Implementing a DPSS-based filter amounts to finding the best fit values $\hat{\mathbfit{x}}$ for the equation
\begin{equation}
    \mathbfit{V} = {\bf A} \mathbfit{x},
\end{equation}
where the vector $\mathbfit{x}$ contains the coefficients of the DPSS fit, and the design matrix ${\bf A}$ has matrix elements
\begin{equation}
    \label{eq:dpss-design-matrix}
    A_{tk} = \phi_k(t) e^{-i2\pi f_{r,0} (t-t_0)},
\end{equation}
where $t_0$ is a reference time and the phase factor $\exp\bigl(-i2\pi f_{r,0} (t-t_0)\bigr)$ shifts the DPSS modes to be centered on the filter center.
The linear least-squares solution for the best-fit DPSS mode coefficients $\hat{\mathbfit{x}}$ is
\begin{equation}
    \label{eq:dpss-best-fit}
    \hat{\mathbfit{x}} = \bigl({\bf A}^\dagger {\bf A} \bigr)^{-1} {\bf A}^\dagger \mathbfit{V}.
\end{equation}
Since the DPSS modes form an orthonormal set, ${\bf A}^\dagger {\bf A}$ is just the identity, and since the filtered visibilities are just the best-fit model for the given set of DPSS modes, the DPSS-filtered visibilities $\mathbfit{V}'$ can be computed via
\begin{equation}
    \label{eq:dpss-filtered-visibilities}
    \mathbfit{V}' = {\bf AA}^\dagger \mathbfit{V}.
\end{equation}
The DPSS filter matrix is therefore
\begin{equation}
    \label{eq:dpss-filter}
    T_{tt'}^{\rm DPSS} = e^{-i2\pi f_{r,0} (t - t')} \sum_{k=0}^N \phi_k (t) \phi_k(t')^*,
\end{equation}
where the sum is truncated to $N < N_t - 1$ terms rather than all of the terms admitted by~\autoref{eq:dpss-eigenvalue-equation}.
Using only a handful of modes rather than the entire set of DPSS modes is required for~\autoref{eq:dpss-filter} to act as a filter, since the complete set of DPSS modes form a basis (and the sum would then converge to the identity).
Using the Fourier Shift Theorem, we can quickly write down the DPSS filter transfer matrix as
\begin{equation}
    \label{eq:dpss-filter-transfer-matrix}
    \bar{T}_{f_r f_r'}^{\rm DPSS} = \sum_k \bar{\phi}_k(f_r - f_{r,0}) \bar{\phi}_k(f_r' - f_{r,0})^*.
\end{equation}

\begin{figure*}
    \includegraphics[width=\textwidth]{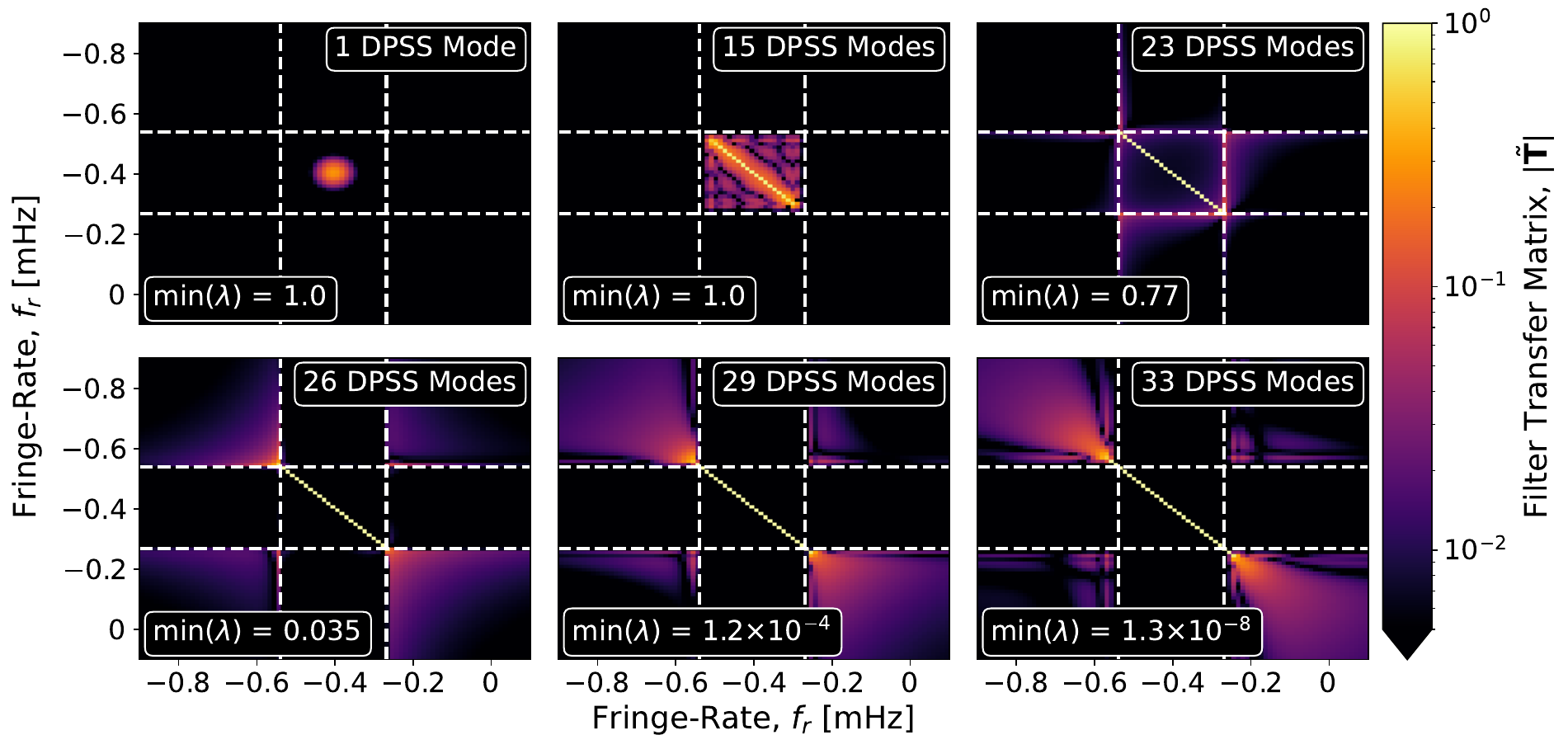}
    \caption{
        Filter transfer matrices for a DPSS-based filter with filter bounds $f_{r,1}$ and $f_{r,2}$ plotted as dashed white lines.
        Each panel shows the filter produced using a different number of DPSS modes.
        Including more modes provides a better representation of the signal within the filter bounds and less covariance between modes within the filter bounds, but comes with the cost of increased power outside the filter bounds.
    }
    \label{fig:dpss-filter-matrices}
\end{figure*}

In~\autoref{fig:dpss-filter-matrices}, we show how the shape of the filter transfer matrix is affected by the number of DPSS modes used to construct the filter.
Since the DPSS modes are sorted by their spectral concentration, setting a cutoff on the number of modes used is equivalent to setting a minimum spectral concentration threshold.
The literature~\citep[e.g.,][]{Slepian:1978,Percival&Walden:1993,Hristopulos:2020} notes that the spectral concentration of DPSS modes follow a bimodal distribution in the sense that most modes have a spectral concentration near unity or near zero, with very few modes occupying intermediate values.
\autoref{fig:dpss-filter-matrices} provides an alternate view of this bimodal behavior: adding more modes provides a more accurate reconstruction of the signal within the filter bounds up to some critical number of modes; including additional modes beyond the critical mode does not significantly change the filter characteristics within the filter bounds, but rather adds additional structure beyond the filter bounds.
Additionally, including too few modes may result in strong correlations within the filter bounds, which may complicate downstream analyses.
Consequently, choosing a permissive eigenvalue cutoff (i.e., setting ${\rm min}(\lambda) \ll 1$) effectively guarantees that the DPSS filter will accurately reconstruct the signal within the filter bounds at the expense of retaining some signal outside the filter bounds; however, care must be taken to not set too small of an eigenvalue cutoff, since this may retain too much power beyond the filter bounds and mitigate the efficacy of the filter.
An investigation of how to design an ``optimal'' DPSS filter that balances signal reconstruction accuracy within the filter bounds against signal leakage outside the filter bounds is beyond the scope of this paper, but~\autoref{fig:dpss-filter-matrices} provides a useful bit of wisdom.
In accordance with the findings from~\citet{Garsden:2024} and~\citet{Bull:2024}, fewer DPSS modes leads to better signal confinement within the filter bounds, while more DPSS modes leads to a more accurate reconstruction of the signal within the filter bounds at the expense of additional leakage beyond the filter bounds.

\subsubsection{Coherent Time Averaging}
Recall that the signal loss formalism presented in Section~\ref{sec:SIGNAL-LOSS-FORMALISM} generally applies to any linear time-based operation acting on a time stream of visibilities.
Many experiments employ fringe-stopped coherent time averages to obtain improved signal-to-noise, and the formalism presented here may be used to compute the signal loss associated with this averaging.
A fringe-stopped coherent time average can be treated as a linear filter with design matrix elements
\begin{equation}
    T_{tt'} = w_{tt'} e^{-i\psi_{tt'}},
\end{equation}
where the fringe-stopping phase $\psi_{tt'}$ is given by~\citep[e.g.,][]{Zhang:2018}
\begin{equation}
    \psi_{tt'} = 2\pi \mathbfit{u}^T (\boldsymbol{\Gamma}_{tt'} - \mathbf{1}) \hat{\mathbfit{n}}_t,
\end{equation}
where $\mathbfit{u} = \mathbfit{b}/\lambda$ is the baseline in units of wavelengths, $\bvec{n}_t$ is the phase center at time $t$, $\mathbf{1}$ is the identity, and $\boldsymbol{\Gamma}_{tt'}$ is a rotation matrix that accounts for the drift in the fringe phase in the $\bvec{n}_t$ direction between times $t$ and $t'$ due to Earth rotation.
The weights $w_{tt'}$ just keep track of the number of integrations that are summed together in each averaged bin, and may be written as
\begin{equation}
    w_{tt'} = \begin{cases}
        1/N_{\rm avg}, \qquad |t - t'| \leq \delta t/2 \\
        0, \qquad\qquad\qquad {\rm else}
    \end{cases},
\end{equation}
where $N_{\rm avg}$ is the number of samples averaged into each bin at time $t$ and $\delta t$ is the duration of the coherent average.
Note that this is not an optimal time averaging implementation, but this fringe-stopped averaging is approximately optimal (Martinot \& Aguirre, in preparation).

\section{Monte-Carlo Simulations}
\label{sec:SIMULATIONS}
In this section, we compare predictions from the $m$-mode based formalism against Monte Carlo estimates using a suite of visibility simulations.
These comparisons use $m$-mode power spectra and simulated visibilities for the Phase II HERA instrument~\citep{Berkhout:2024}.
All of the calculations performed in this section are computed on a per-baseline basis; however, since we are not considering covariances between baselines, we will omit baseline labels (i.e., it should be understood that each baseline has its own $m$-mode power spectrum, the fringe-rate filter for one baseline will typically not be the same as that of another baseline, and so on).

The Monte Carlo is performed over multiple realizations of a sky-locked Gaussian random field with a flat power spectrum $P(k) = {\rm constant}$.
The visibility simulations were performed using the \texttt{fftvis}\footnote{\url{https://github.com/tyler-a-cox/fftvis}} package, which evaluates the measurement equation by using the \texttt{finufft}\footnote{\url{https://github.com/flatironinstitute/finufft}}~\citep{finufft} package to perform a non-uniform Fast Fourier Transform of the beam-weighted sky.
The specific intensity in each pixel of the simulated sky map is Gaussian distributed and independent across different frequencies, with variance
\begin{equation}
    \sigma^2(\nu) = \sigma^2_0 \frac{\Delta V(\nu_0)}{\Delta V(\nu)},
\end{equation}
where $\sigma^2_0$ is the variance at reference frequency $\nu_0$, and $\Delta V(\nu)$ is the differential comoving volume for the given pixel size and channel width.
Both the visibility simulations and the $m$-mode spectrum calculations used a CST-simulated model of the Phase II HERA beam~\citep{Fagnoni:2021}.
The visibility simulations cover only a small subset of the HERA array: the simulations contain 14 antennas, with a maximum East-West projected baseline length of 88 meters, and a maximum North-South projected baseline length of 25 meters.
None of the systematics known to be in HERA data (e.g., RFI, mutual coupling, and calibration errors) are included in the simulated visibilities---we defer investigations of how various known systematics couple to fringe-rate filtering effects to future work.
A summary of additional simulation parameter information can be found in~\autoref{table:simulation-details}.

\begin{table}
    \begin{tabular}{cc}
        \hline
        Parameter & Value \\
        \hline
        Number of Realizations & 49 \\
        Number of Times & 1000 \\
        Integration Time & 86.16 s \\
        LST Coverage & 24 hr \\
        Number of Frequencies & 512 \\
        Channel Width & 351.56 kHz \\
        Frequency Range & 60 \--- 240 MHz \\
        Input $P(k)$ & Flat Spectrum \\
        $P(k)$ at 150 MHz & 23,357 mK$^2$ $(h^{-1} {\rm Mpc})^3$ \\
        Sky $N_{\rm side}$ & 128 \\
        \hline
    \end{tabular}
    \caption{
        Monte Carlo visibility simulation parameters.
    }
    \label{table:simulation-details}
\end{table}

\subsection{Fringe-Rate Covariance}
We begin by comparing the Monte Carlo estimate of the fringe-rate covariance against the analytic expectation at a single frequency.
Since numerical estimates of covariance matrices tend to slowly converge~\citep{Dodelson&Schneider:2013,Taylor&Joachimi:2014,Cheng:2018}, we do not directly compare the Monte Carlo estimate of the fringe-rate covariance against the analytic covariance.
We instead perform an additional Monte Carlo over $m$-mode realizations and compare the fringe-rate covariance estimated with the ``$m$-mode Monte Carlo'' against the fringe-rate covariance estimated with the suite of visibility simulations.
The $m$-mode Monte Carlo is performed by computing
\begin{equation}
    \label{eq:m-mode-monte-carlo}
    V^r(\nu,t) = \sum_m \sqrt{M_{m\nu}} V_m^r e^{-im\omega_\oplus t},
\end{equation}
where $M_{m\nu}$ is the $m$-mode power spectrum at frequency $\nu$, and the $m$-modes are Gaussian distributed so that $V_m^r \sim \mathcal{N}(0, \sigma^2)$.
For each set of simulated visibilities we estimate the fringe-rate covariance $\hat{C}_\nu(f_r, f_r')$ via
\begin{equation}
    \label{eq:monte-carlo-covariance}
    \hat{C}_\nu(f_r, f_r') = \frac{1}{N_r} \sum_r \bar{V}^r(\nu,f_r) \bar{V}^r(\nu,f_r')^*,
\end{equation}
where $N_r$ is the number of realizations and $\bar{V}^r(\nu, f_r)$ is the fringe-rate transform of~\autoref{eq:m-mode-monte-carlo}.

\begin{figure*}
    \includegraphics[width=\textwidth]{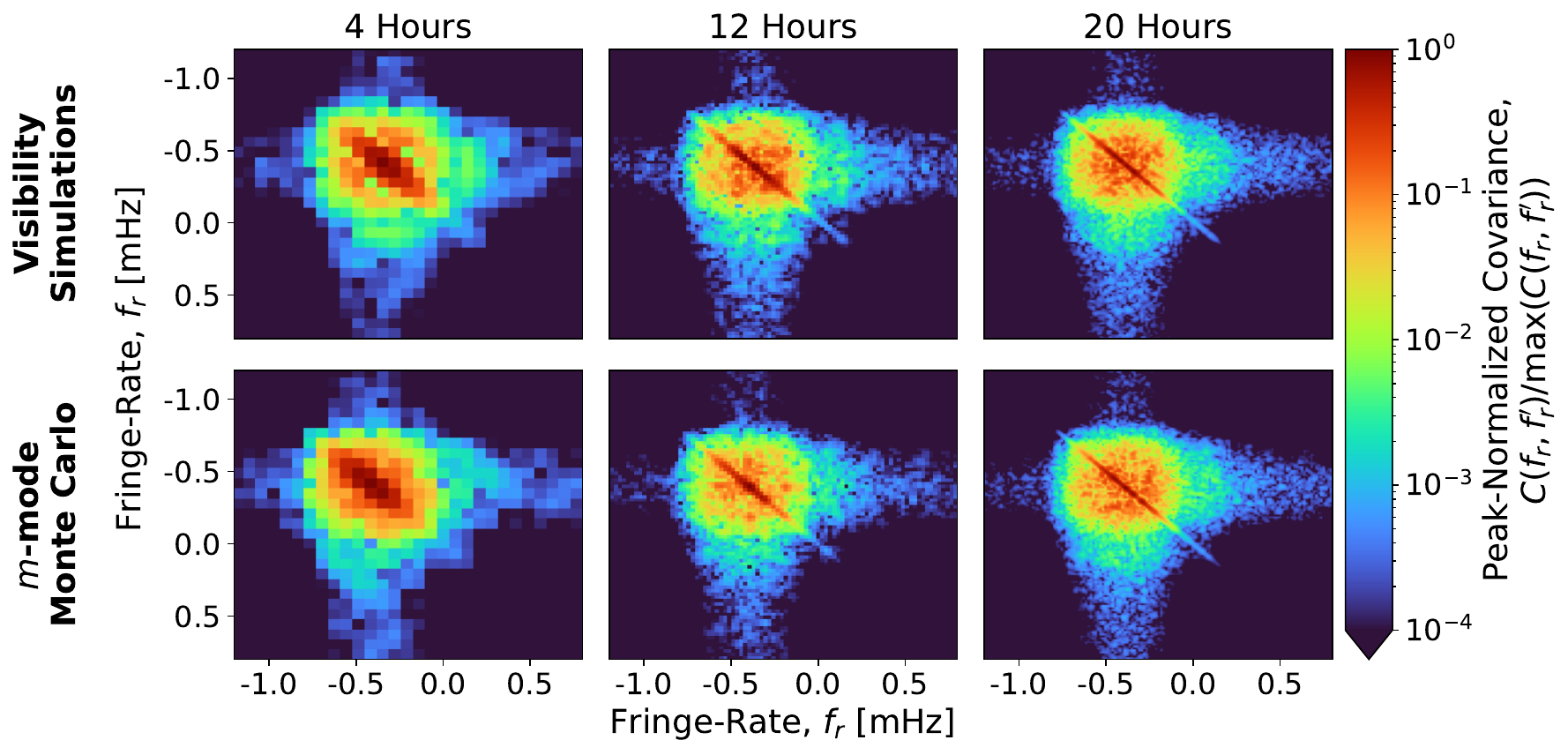}
    \caption{
        Comparison of the fringe-rate covariance estimated from a Monte Carlo over sky realizations (top row) against a Monte Carlo over $m$-mode power spectrum realizations (bottom row).
        The comparison is made for a 14.6-meter East-West baseline at 130 MHz, with the covariance matrices estimated using 49 realizations of the underlying signal.
        Each column shows the fringe-rate covariance for a different length of observing time, with a Hann window used when taking the fringe-rate transform.
        Although~\autoref{fig:fringe-rate-cov} shows that the fringe-rate covariance converges to a diagonal matrix in the limit of a full sidereal day of observation, the Monte Carlo estimates do not feature this behavior---since covariance matrix estimates are slow to converge, sample variance effects create a substantial amount of off-diagonal structure.
        Since the fringe-rate transforms were taken with a Hann window, there is very little additional sidelobe structure coming from the non-periodic boundary, so the main differences between columns are the resolution and small-scale features in the Monte Carlo noise.
    }
    \label{fig:frate-cov-comparisons}
\end{figure*}

In~\autoref{fig:frate-cov-comparisons}, we compare the two Monte Carlo estimates of the fringe-rate covariance for three different total observation times.
While there are differences in the detailed structures due to sample variance effects, the shapes of the fringe-rate covariance matrices are very similar---both approaches produce strong diagonal structures, and the amplitude of off-diagonal structures are similar between the two approaches.
As an additional sanity check, we ran the $m$-mode Monte Carlo for a very large number of realizations and found that the covariance matrix estimates converged to the analytic expectation (i.e.,~\autoref{fig:fringe-rate-cov}).
This result builds confidence in the idea that Monte Carlo signal loss estimates will converge to the expected signal loss.

\subsection{Signal Loss from Fringe-Rate Filtering}
\label{sec:MC-SIGNAL-LOSS}
In this section, we compare Monte Carlo signal loss estimates against the analytic expectation from the $m$-mode based formalism.
As our test case, we will consider signal loss associated with the ``main lobe'' fringe-rate filter, and we begin in Section~\ref{sec:FRF-IMPLEMENTATION} with a discussion of how such a filter may be designed.
In Section~\ref{sec:HERA-FORECAST}, we compute the expected signal loss for the main lobe filter described in Section~\ref{sec:FRF-IMPLEMENTATION} applied to the HERA Phase II instrument, the results of which are summarized in~\autoref{fig:h4c-signal-loss-estimates}.
In Section~\ref{sec:POWER-SPECTRUM-LOSSES}, we review the delay spectrum approach for estimating power spectra, describe how the visibility simulations are used to estimate signal loss, and discuss the results of the signal loss comparison summarized in~\autoref{fig:signal-loss-comparison}.

\subsubsection{Fringe-Rate Filter Implementation}
\label{sec:FRF-IMPLEMENTATION}
The main lobe fringe-rate filter is a signal processing tool designed to reject systematic features in the data that are inconsistent with the expected fringe-rate structure of the cosmological 21-cm signal.
The conceptual design of the filter is fairly straightforward: the filter is a top-hat centered on the peak of the fringe-rate profile, with the width set to retain a user-defined fraction of power in the cosmological signal.
The filter is thus characterized by the filter bounds $f_{r,1}, f_{r,2}$.
When implemented as a Fourier filter, where $\bar{V}'(f_r) = \bar{T}(f_r) \bar{V}(f_r)$, the main lobe filter $\bar{T}(f_r; f_{r,1}, f_{r,2})$ takes the form
\begin{equation}
    \tilde{T}(f_r;f_{r,1},f_{r,2}) = \begin{cases}
        1, \qquad f_{r,1} \leq f_r \leq f_{r,2} \\
        0, \qquad {\rm else}
    \end{cases}.
\end{equation}
The filter bounds may be tuned to a desired level of signal loss $L$ by choosing the filter bounds so that the integrated power in the fringe-rate profile outside of the filter bounds is equal to the desired signal loss.

We may use the tools from Section~\ref{sec:FRATE-PROFILES} to determine the filter bounds $f_{r,1}$ and $f_{r,2}$.
One approach is to first compute the cumulative fringe-rate distribution $\mathcal{F}(f_r)$ via
\begin{equation}
    \label{eq:cumulative-frate-distribution}
    \mathcal{F}(f_r) \equiv \frac{\int_{-\infty}^{f_r} P(f_r') df_r'}{\int_\mathbb{R} P(f_r') df_r'},
\end{equation}
where $P(f_r)$ is the fringe-rate profile (\autoref{eq:fringe-rate-profile}).
Following this, percentile cuts $p_1$ and $p_2$ are chosen so that $L = 1 - (p_2 - p_1)$, and these percentile cuts can be used to compute the filter bounds from the cumulative fringe-rate distribution with
\begin{align}
    \nonumber
    \mathcal{F}(f_{r,1}) &= p_1, \\
    \mathcal{F}(f_{r,2}) &= p_2.
    \label{eq:main-lobe-filter-bounds}
\end{align}
By construction, a top-hat Fourier filter with these filter bounds will incur the desired level of signal loss.
Note that this approach to determining the filter bounds is guaranteed to produce a fringe-rate filter that is centered on the peak fringe-rate for symmetric fringe-rate profiles---for applications where the fringe-rate profiles are significantly skewed, however, it would be more appropriate to instead compute the cumulative fringe-rate distribution by integrating out from the peak fringe-rate.

The approach for designing a main lobe filter for data that spans a range of frequencies, or \emph{spectral window}, is somewhat more complicated and requires an additional choice from the analyst.
Since the fringe-rate profiles are frequency dependent, a straightforward approach is to compute the cumulative fringe-rate distribution at each frequency $\mathcal{F}(\nu,f_r)$ using the frequency-dependent fringe-rate profiles $P(\nu,f_r)$.
This approach would obtain a different fringe-rate filter for each frequency, and therefore carries the risk of injecting additional spectral structure into the filtered data.
Since spectral smoothness is a chief concern in the context of 21-cm cosmology, we instead opt to first compute an ``effective'' fringe-rate profile by taking a weighted average of the frequency-dependent fringe-rate profiles across the spectral window.
The effective fringe-rate profile $P^{\rm eff}(f_r;\nu_1, \nu_2, B)$ is computed as
\begin{equation}
    \label{eq:effective-frate-profile}
    P^{\rm eff}(f_r;\nu_1, \nu_2, B) \equiv \frac{\int_{\nu_1}^{\nu_2} B(\nu)^2 P(\nu, f_r) d\nu}{\int_{\nu_1}^{\nu_2} B(\nu)^2 d\nu},
\end{equation}
where $\nu_1$ and $\nu_2$ are the bounds of the spectral window and $B(\nu)$ is the taper used when computing the delay transform (\autoref{eq:delay-transform}) for power spectrum estimation.
We then use the effective fringe-rate profile to compute the cumulative fringe-rate distribution and filter bounds according to~\autoref{eq:cumulative-frate-distribution} and~\autoref{eq:main-lobe-filter-bounds}.
The fringe-rate filters designed with this approach are constant over the spectral window and the filters therefore do not inject additional spectral structure into the data.

Although the filter is designed to meet a specific signal loss requirement when implemented as a Fourier filter, we choose to use a DPSS-based filter as described in Section~\ref{sec:DPSS-FILTERS}, since this is the same type of filter that HERA uses to mitigate mutual coupling.
Rather than use a fixed number of DPSS modes to construct the fringe-rate filters, we set the number of DPSS modes by enforcing a spectral concentration cutoff $\lambda_{\rm cut}$, such that only DPSS modes $\phi_k(t)$ with $\lambda_k \geq \lambda_{\rm cut}$ are used for filtering.
Given the filter half-width $\delta f_r = |f_{r,2} - f_{r,1}|/2$, the number of integrations, and the integration time, the DPSS modes and spectral concentrations are computed using \texttt{scipy}~\citep{scipy}.
For our particular test case, we compute the filter bounds with percentile cuts $p_1, p_2$ of 5\% and 95\%, respectively, keep all DPSS modes with a spectral concentration greater than $10^{-9}$, and use a Blackman-Harris taper~\citep{Blackman:1958} for computing the effective fringe-rate profiles---these choices are summarized in~\autoref{table:filter-params}.
Our choice of spectral concentration cutoff results in ``permissive'' fringe-rate filters, since the modes with low spectral concentrations retain power slightly beyond the filter edges.
Consequently, we ought to expect the actual signal loss induced by the filter to be somewhat less than the intended signal loss.

\begin{table}
    \begin{tabular}{cc}
        \hline
        Parameter & Value \\
        \hline
        $p_1$ & 0.05 \\
        $p_2$ & 0.95 \\
        $\lambda_{\rm cut}$ & $10^{-9}$ \\
        $B(\nu)$ & Blackman-Harris \\
        \hline
    \end{tabular}
    \caption{
        Fringe-rate filter design parameters.
    }
    \label{table:filter-params}
\end{table}

\subsubsection{Signal Loss Projections for HERA}
\label{sec:HERA-FORECAST}
We apply our signal loss formalism to forecast the expected signal loss when applying the main lobe filter from Section~\ref{sec:FRF-IMPLEMENTATION} to HERA data.
We compute the signal loss for all of the baselines and all of the spectral windows above the FM band that will be used in an upcoming analysis.
When designing the main lobe filter, we use the percentile cuts, eigenvalue cutoff, and frequency taper listed in~\autoref{table:filter-params}.
We use~\autoref{eq:signal-loss-general-basis} with the covariance and filters expressed in the fringe-rate domain to compute the signal loss for each baseline.
Since we are computing the signal loss per spectral window instead of per frequency, we replace the covariance $\mathbf{C}$ with the effective covariance $\mathbf{C}^{\rm eff}$ obtained by taking a weighted average of the per-frequency fringe-rate covariance matrices via
\begin{equation}
    \label{eq:effective-covariance}
    C^{\rm eff} (f_r, f_r') = \frac{\sum_\nu B(\nu)^2 C_\nu(f_r, f_r')}{\sum_\nu B(\nu)^2},
\end{equation}
where the sums are taken over all frequencies within a given spectral window.

\begin{figure*}
    \includegraphics[width=\textwidth]{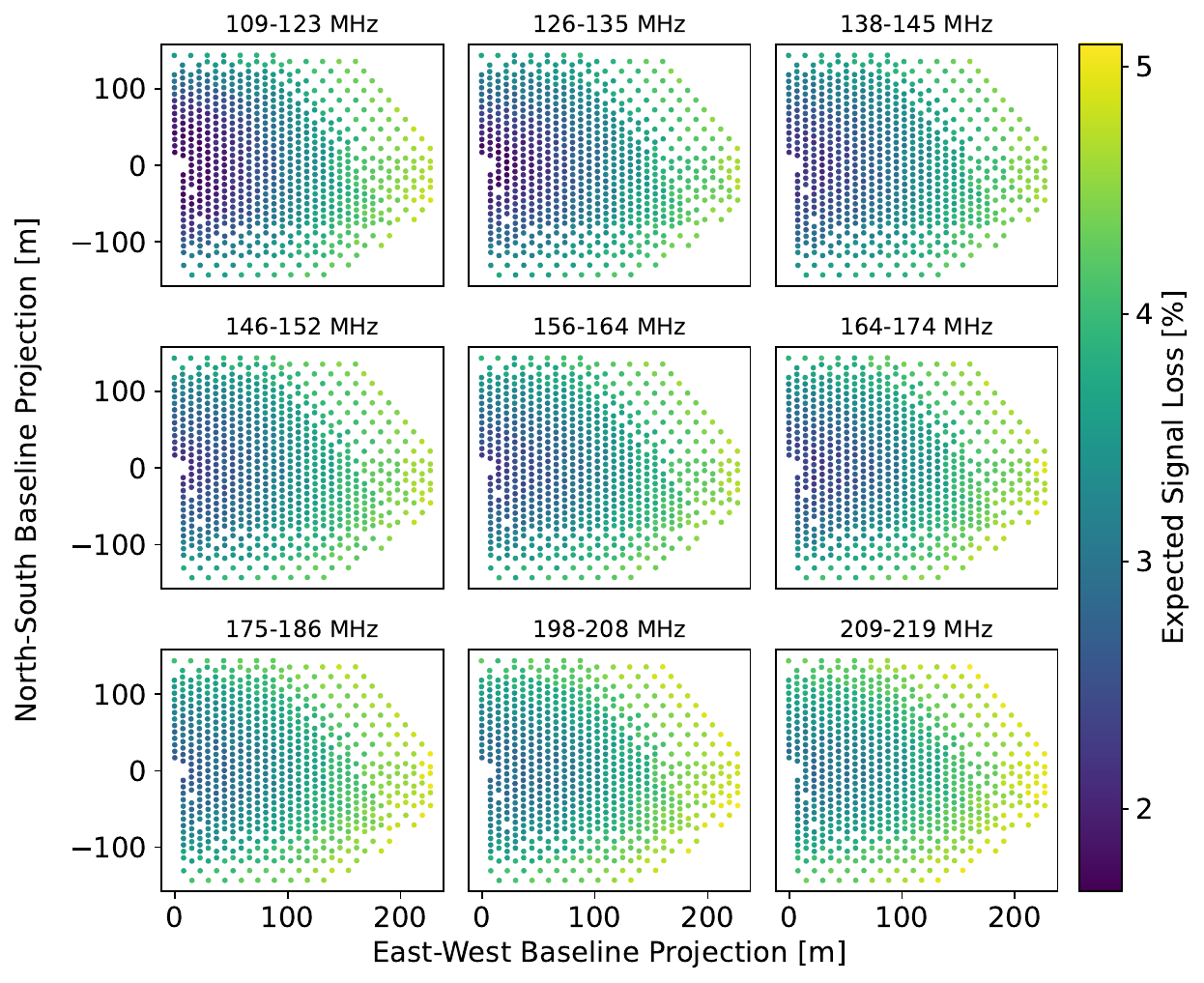}
    \caption{
        Signal loss estimated using the $m$-mode formalism, calculated for the spectral windows used in an upcoming HERA power spectrum analysis.
        In each panel, the horizontal axis corresponds to the East-West projected baseline length, while the vertical axis corresponds to the North-South projected baseline length.
        The color of each marker gives the estimated signal loss incurred by the main lobe filter described in Section~\ref{sec:FRF-IMPLEMENTATION}.
        The filters were designed to retain 90\% of the cosmological signal; all of the baselines show less loss than this specification due to the permissive nature of the DPSS filters employed, as discussed in Section~\ref{sec:FRF-IMPLEMENTATION}.
    }
    \label{fig:h4c-signal-loss-estimates}
\end{figure*}

We show the predicted signal loss for each baseline and spectral window in~\autoref{fig:h4c-signal-loss-estimates}.
Even though the filters were designed to retain 90\% of the cosmological signal, we find that the signal loss tends to hover around a few percent and increases with projected East-West baseline length.
The fact that the filters incur less signal loss than the design specification is expected, since we used a ``permissive'' eigenvalue cutoff---recall from Section~\ref{sec:DPSS-FILTERS} that this choice of eigenvalue cutoff results in high fidelity reconstruction of the signal within the filter bounds at the expense of additional leakage beyond the filter bounds.
We conclude that the method for designing fringe-rate filters described in Section~\ref{sec:FRF-IMPLEMENTATION} will not incur signal loss beyond the desired threshold used to construct the filter.

\subsubsection{Power Spectra and Signal Loss}
\label{sec:POWER-SPECTRUM-LOSSES}
We numerically estimate the signal loss incurred by the main lobe fringe-rate filter by forming delay spectra of the simulated visibilities and comparing the average power in the delay spectra before and after filtering.
The delay spectrum~\citep{Parsons:2012b} is a method for estimating the cosmological power spectrum $\mathcal{P}(k_\parallel, \mathbfit{k}_\perp)$ directly from the delay-transformed visibilities $\tilde{V}(\tau)$ (\autoref{eq:delay-transform}) through
\begin{equation}
    \label{eq:delay-spectrum-estimator}
    \mathcal{P}(k_\parallel, \mathbf{k}_\perp) = \mathcal{A}\big| \tilde{V}(\tau)\big|^2,
\end{equation}
where $\mathcal{A}$ is a normalization factor that depends on the choice of cosmology and observing parameters.
In the delay spectrum method, each baseline samples a different cosmological Fourier mode in the plane of the sky (i.e., $\mathbfit{b} \propto \mathbfit{k}_\perp$), while each delay samples a different cosmological Fourier mode along the line of sight (i.e., $\tau \propto k_\parallel$).
In a typical HERA power spectrum analysis~\citep{HERA:2022b,HERA:2023}\footnote{The actual analyses employ an interleaved incoherent time average, where pairs of adjacent times are cross-multiplied prior to averaging in order to avoid a noise bias. Since our simulations are noiseless, we omit this step; however, our signal loss method may be adapted to interleaved averages.}, the per-baseline power spectra are computed by incoherently averaging power spectrum estimates across time, so the delay spectrum estimate $\hat{\mathcal{P}}(\tau)$ can be written as
\begin{equation}
    \label{eq:delay-spectrum}
    \hat{\mathcal{P}}(\tau) = \frac{\mathcal{A}}{N_t} \sum_t \big| \tilde{V}(\tau,t)\big|^2,
\end{equation}
where $N_t$ is the number of times included in the average.

Following the framework of Section~\ref{sec:SIGNAL-LOSS-FORMALISM}, we estimate the signal loss by comparing the average filtered power against the average unfiltered power.
To estimate this from the visibility simulations, we compute~\autoref{eq:delay-spectrum} for each realization $r$, average the result over all delays and realizations, then take the ratio of the computed result between the filtered visibilities and the unfiltered visibilities.
The signal loss estimate $\hat{L}$ is thus
\begin{equation}
    \label{eq:signal-loss-estimate}
    \hat{L} = 1 - \frac{\sum_{r,\tau} \hat{\mathcal{P}}^r_{\rm filt}(\tau)}{\sum_{r,\tau} \hat{\mathcal{P}}^r_{\rm unfilt}(\tau)}.
\end{equation}
In the appropriate limit, this signal loss estimate will converge to the analytic expectation (\autoref{eq:expected-signal-loss}), as we show below.
In Appendix~\ref{sec:SIGNAL-LOSS-DISTRIBUTIONS}, we explore the statistical properties of the Monte Carlo signal loss estimate $\hat{L}$ and provide an alternate proof that the signal loss estimator~\autoref{eq:signal-loss-estimate} converges to the analytic expectation with enough sky realizations.

Since the data is discrete, we begin by writing the discrete version of the delay transformed visibilities, which is
\begin{equation}
    \label{eq:discrete-delay-transform}
    \tilde{V}^r(\tau,t) = \sum_\nu B(\nu) V^r(\nu,t) e^{-i2\pi\nu\tau},
\end{equation}
where the sum is taken over the spectral window from $\nu_1$ to $\nu_2$.
Similarly, the delay transform of the fringe-rate filtered visibilities is
\begin{equation}
    \label{eq:delay-transformed-filtered-visibilities}
    \tilde{V}^r_{\rm filt}(\tau, t) = \sum_{\nu,t'} B(\nu) T_{tt'} V^r(\tau,t') e^{-i2\pi\nu\tau},
\end{equation}
where the sum over $t'$ is taken across all times.
Plugging~\autoref{eq:discrete-delay-transform} and~\autoref{eq:delay-transformed-filtered-visibilities} into the signal loss estimator~\autoref{eq:signal-loss-estimate}, we get
\begin{equation}
    \label{eq:signal-loss-estimate-from-visibilities}
    \hat{L} = 1 - \frac{\sum_{r,\tau,t} \big| \sum_{\nu,t'} B(\nu) T_{tt'} V^r(\nu, t') e^{-i2\pi\nu\tau}\big|^2}{\sum_{r,\tau,t} \big|\sum_\nu B(\nu) V^r(\nu, t) e^{-i2\pi\nu\tau}\big|^2}.
\end{equation}
In both the numerator and denominator of~\autoref{eq:signal-loss-estimate-from-visibilities}, the sum over delay modes acts only on the complex exponential terms.
For a large enough spectral window and an appropriate choice of taper, the sum may be approximated as
\begin{equation}
    \label{eq:delay-sidelobe-approx}
    \sum_\tau e^{-i2\pi(\nu-\nu')\tau} \approx N_\tau \delta_{\nu\nu'},
\end{equation}
where $N_\tau$ is the number of delay modes (which is the same as the number of frequencies in the spectral window) and $\delta_{\nu\nu'}$ is the Kronecker delta.
With this approximation, the signal loss estimate reduces to
\begin{equation}
    \label{eq:simplified-signal-loss-estimate}
    \hat{L} = 1 - \frac{\sum_{r,t,\nu} B(\nu)^2 \sum_{t',t''} T_{tt'} V^r(\nu,t') V^r(\nu,t'')^* T_{tt''}^*}{\sum_{r,t,\nu} B(\nu)^2 |V^r(\nu,t)|^2}.
\end{equation}
Evaluating the average over realizations yields an estimate of the time-time covariance $\hat{\mathbf{C}}$, and the weighted average over frequency converts this into an estimate of the effective time-time covariance $\hat{\mathbf{C}}^{\rm eff}$ (c.f.~\autoref{eq:effective-covariance}).
The remaining sums over time can be rewritten as matrix products and traces, which ultimately allows us to write the signal loss estimate as
\begin{equation}
    \label{eq:signal-loss-estimate-final}
    \hat{L} = 1 - \frac{{\rm tr}\bigl( {\bf T} \hat{\bf C}^{\rm eff} {\bf T}^\dagger \bigr)}{{\rm tr} \hat{\bf C}^{\rm eff}}.
\end{equation}
In the limit of an infinite number of realizations, the covariance estimate converges to the expected covariance, and therefore the Monte Carlo signal loss estimate converges to the analytic expectation computed with the effective time-time (or fringe-rate) covariance.

\begin{figure*}
    \includegraphics[width=\textwidth]{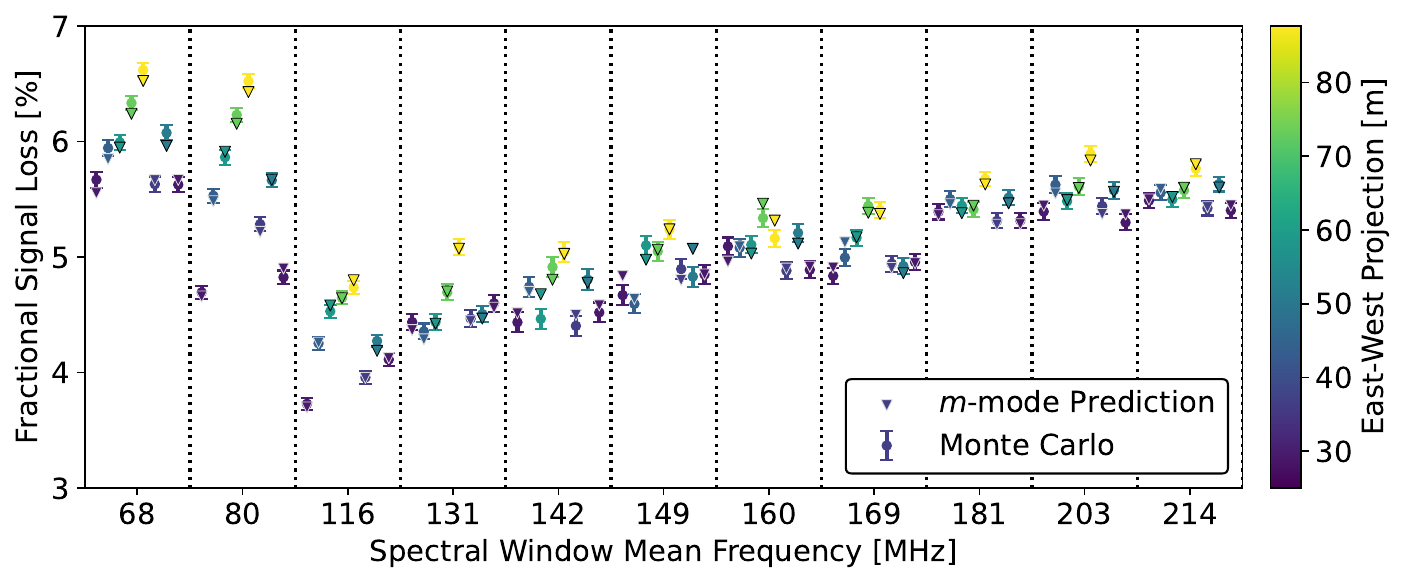}
    \caption{
        Comparison of the two signal loss estimation routines described in Section~\ref{sec:POWER-SPECTRUM-LOSSES}.
        The triangles denote the expected signal loss computed according to~\autoref{eq:signal-loss-general-basis} in the fringe-rate domain with the effective fringe-rate covariance.
        The circles denote the signal loss estimated using the visibility simulations described in Section~\ref{sec:SIMULATIONS}, and the associated error bars denote the 68\% confidence intervals computed from the expected distribution of Monte Carlo signal loss estimates derived in Appendix~\ref{sec:SIGNAL-LOSS-DISTRIBUTIONS}.
        The vertical dashed lines mark the boundaries between different spectral windows, and all of the markers between adjacent pairs of dashed lines correspond to the same spectral window (i.e., horizontal space is added between signal loss values for different baselines to make the figure more readable).
    }
    \label{fig:signal-loss-comparison}
\end{figure*}

In~\autoref{fig:signal-loss-comparison} we compare the Monte Carlo signal loss estimates against the analytic expectation for all of the baselines in the simulation with a projected East-West length longer than 28 meters in 11 different spectral windows.
The spectral windows are each roughly 10 MHz wide and span the full extent of the HERA Phase II bandwidth, with the lowest spectral window centered on 68 MHz and the highest spectral window centered on 214 MHz.
The Monte Carlo signal loss estimates are plotted with error bars denoting the 68\% confidence intervals; the confidence intervals were computed using the expected signal loss distributions derived in Appendix~\ref{sec:SIGNAL-LOSS-DISTRIBUTIONS}.
We find that the Monte Carlo signal loss estimates are in excellent agreement with the analytic expectation across the board.
We therefore conclude that the signal loss formalism presented in this paper is consistent with a more traditional Monte Carlo approach.

\section{Conclusion}
\label{sec:CONCLUSION}
The introduction of fringe-rate filtering as a commonly employed analysis tool in interferometric 21-cm cosmology experiments brought with it a demand for an improved understanding of how to characterize the signal loss properties of these filters.
In this paper, we proposed an analytic framework for predicting the signal loss properties of fringe-rate filters based on tools developed for the $m$-mode analysis of visibilities.
Using the formalism presented in Martinot \& Aguirre (in preparation), we showed in Section~\ref{sec:TIME-EVOLUTION} how both the time-time covariance and fringe-rate covariance are uniquely determined by the instrumental $m$-mode power spectrum (\autoref{eq:m-mode-power-spectrum}) and a weighting matrix that encodes details of the observation and any tapering that is applied when taking the fringe-rate transform of the data.
This formalism provides a necessary extension beyond the approximate treatment of~\citet{Parsons:2016}, since we showed in Section~\ref{sec:PREV-WORK} that the instantaneous fringe-rate model cannot correctly characterize the fringe-rate response of short baselines.
We adapted the framework from Martinot \& Aguirre (in preparation) to devise a method of analytically computing the signal loss for linear time-based operations applied to drift-scanning interferometric data, which we validated against a Monte Carlo signal loss analysis using HERA as a test case.
In analyzing the Monte Carlo signal loss estimates, we developed a statistically rigorous characterization of the errors in the Monte Carlo analysis---these results provide a firm footing for other Monte Carlo signal loss analyses, and the details of our statistical analysis are reviewed in Appendix~\ref{sec:SIGNAL-LOSS-DISTRIBUTIONS}.

We conclude that our $m$-mode based signal loss formalism may be used to quickly and accurately predict the signal loss incurred by applying temporal filters to drift-scanning interferometric data.
Since all of our calculations are analytic, our approach offers substantial computational gains over traditional Monte Carlo analyses and provides exact signal loss predictions.
Our signal loss formalism may therefore be used as a cheaper, more accurate alternative to a Monte Carlo signal loss analysis.
Additionally, the agreement of our mock analysis pipeline with our analytic predictions bolsters validation efforts that are crucial to the success of 21-cm cosmology experiments~\citep{Aguirre:2022}.

\section*{Acknowledgements}
The authors thank Daniel Jacobs, Bryna Hazelton, Miguel Morales, Aaron Ewall-Wice, and Aaron Parsons for helpful discussions.
AL and RP acknowledge support from the Trottier Space Institute, the Canadian Institute for Advanced Research (CIFAR) Azrieli Global Scholars program, a Natural Sciences and Engineering Research Council of Canada (NSERC) Discovery Grant, a NSERC/Fonds de recherche du Québec -Nature et Technologies NOVA grant, the Sloan Research Fellowship, and the William Dawson Scholarship at McGill.
NK acknowledges support from NASA through the NASA Hubble Fellowship grant \#HST-HF2-51533.001-A awarded by the Space Telescope Science Institute, which is operated by the Association of Universities for Research in Astronomy, Incorporated, under NASA contract NAS5-26555.
This work was funded in part by the Canada 150 Research Chairs Program.

\section*{Data Availability}
The simulated data products may be made available upon reasonable request to the corresponding author.

%


\software{
    numpy~\citep{numpy},
    scipy~\citep{scipy},
    astropy~\citep{astropy},
    pyuvdata~\citep{pyuvdata},
    fftvis (\url{https://github.com/tyler-a-cox/fftvis})
}




\bibliography{references}{}
\bibliographystyle{aasjournal}


\appendix

\section{Monte-Carlo Signal Loss Distribution}
\label{sec:SIGNAL-LOSS-DISTRIBUTIONS}
In Section~\ref{sec:SIMULATIONS}, we compared the signal loss predictions from our $m$-mode based formalism against estimates made through a Monte Carlo analysis.
In this appendix, we analytically compute the probability density for the Monte Carlo signal loss estimates under the assumption that the Monte Carlo included enough samples for the average power estimates (the numerator and denominator in~\autoref{eq:signal-loss-estimate-final}, or equivalently the delay average of~\autoref{eq:delay-spectrum} for the filtered and unfiltered visibilities) to Gaussianize.
We additionally show how to compute the variance in the filtered and unfiltered average power estimates, as well as the covariance between the estimates.

\citet{Tan:2021} showed that the errors on time-averaged power spectrum estimates tend to quickly Gaussianize, so we will assume that the Central Limit Theorem (CLT) is applicable for the average power estimates obtained in computing the Monte Carlo signal loss distribution.
We will denote the filtered and unfiltered average power estimates computed with $N$ sky signal realizations as $X = N^{-1} \sum_{r,\tau} \mathcal{P}_{\rm filt}^r(\tau)$ and $Y = N^{-1} \sum_{r,\tau} \mathcal{P}_{\rm unfilt}^r(\tau)$, respectively.
Formally, we will treat these average power estimates as correlated random variables with covariance
\begin{equation}
    \label{eq:X-Y-covariance}
    \mathbf{C} =
    \begin{pmatrix}
        a & c \\
        c & b \\
    \end{pmatrix},
\end{equation}
where $a = {\rm Var}(X)$ and $b = {\rm Var}(Y)$ are the variances in the average power estimates, and $c = {\rm Cov}(X,Y)$ is the covariance between the filtered and unfiltered average power estimates.
Note that since the power spectrum estimates are formed only using auto-baseline pairs without time-interleaving, the covariance matrix $\mathbf{C}$ is real-valued.

In terms of the random variables $X$ and $Y$, the signal loss estimate (\autoref{eq:signal-loss-estimate}) is just $\hat{\mathcal{L}} = 1 - X/Y$.
We can therefore compute the distribution of the signal loss estimates by determining the distribution of the ratio $Z = X/Y$, which is obtained by evaluating
\begin{equation}
    \label{eq:ratio-distribution-integral}
    p_Z(Z) = \int_{-\infty}^\infty p_{XY}(X=Zt, Y=t) |t| {\rm d}t,
\end{equation}
where $p_Z(Z)$ is the distribution of the ratio of average power estimates and $p_{XY}(X,Y)$ is the joint distribution of the filtered and unfiltered average power estimates.
Applying the CLT to the joint distribution gives
\begin{equation}
    p_{XY}(X,Y) = \exp \biggl(-\frac{1}{2} \Bigl( \mathbfit{X} - \boldsymbol{\mu} \Bigr)^T \mathbf{C}^{-1} \Bigl( \mathbfit{X} - \boldsymbol{\mu} \Bigr) \biggr),
\end{equation}
where the filtered and unfiltered average power estimates $X, Y$ have been bundled into a vector $\mathbfit{X}$, and $\boldsymbol{\mu} = \langle \mathbfit{X} \rangle$.
Evaluating this integral gives
\begin{equation}
    \label{eq:ratio-distribution}
    p_Z(Z) \propto \frac{e^{-\gamma/2}}{\alpha} \Bigl[ 1 + \sqrt{\pi} \delta e^{\delta^2} {\rm erf}(\delta)\Bigr],
\end{equation}
where $\alpha, \delta$ are functions of the ratio $Z$, $\gamma$ is a constant that depends on the variance, covariance, and mean of the average power estimates, and ${\rm erf}(\cdot)$ is the error function.
If we define $\mathbfit{q}^T = (Z, 1)$, then the auxiliary parameters are defined as
\begin{align}
    \label{eq:alpha}
    \alpha &= \mathbfit{q}^T \mathbf{C}^{-1} \mathbfit{q}, \\
    \label{eq:beta}
    \beta &= -2\mathbfit{q}^T \mathbf{C}^{-1} \boldsymbol{\mu}, \\
    \label{eq:gamma}
    \gamma &= \boldsymbol{\mu}^T \mathbf{C}^{-1} \boldsymbol{\mu}, \\
    \label{eq:delta}
    \delta &= \frac{\beta}{2\sqrt{2\alpha}}.
\end{align}
Note that if $\beta = 0$, then the second term in~\autoref{eq:ratio-distribution} vanishes and $Z$ becomes Cauchy distributed, so the mean and variance in the signal loss both become undefined.
If we were to find ourselves in this position, then a Monte-Carlo based estimation of the signal loss would be unsuccessful; however, we are firmly not in this regime, since the filtered power spectrum strongly depends on the unfiltered power spectrum.

For the filtering cases explored in this paper, $\gamma/2$ tends to be around a few thousand, so the first term in~\autoref{eq:ratio-distribution} is exponentially suppressed and the distribution of $Z$ is approximately
\begin{equation}
    \label{eq:approx-ratio-distribution}
    p_Z(Z) \propto \frac{\delta}{\alpha} e^{\delta^2 - \gamma/2}{\rm erf}(\delta).
\end{equation}
Additionally, $\delta^2 \leq \gamma/2$ for most values of $Z$ except for a small range of values near $\langle X \rangle / \langle Y \rangle$.
Around these values the ratio $\delta / \alpha$ is roughly constant (and negative), and ${\rm erf}(\delta) \approx -1$, so we may gain some traction analytically by treating the argument of the exponential as the only variable term in~\autoref{eq:approx-ratio-distribution}.
Expanding out the terms in the argument of the exponential and inserting their definitions in terms of ${\bf q}, \boldsymbol{\mu},$ and ${\bf C}$ gives
\begin{equation}
    \label{eq:exponential-argument}
    \delta^2 - \frac{\gamma}{2} = \frac{\bigl({\bf q}^T {\bf C}^{-1} \boldsymbol{\mu}\bigr)^2 - \bigl({\bf q}^T {\bf C}^{-1} {\bf q}\bigr) \bigl(\boldsymbol{\mu}^T {\bf C}^{-1} \boldsymbol{\mu}\bigr)}{2 {\bf q}^T {\bf C}^{-1} {\bf q}}.
\end{equation}
Making the approximations described above, evaluating the matrix products in~\autoref{eq:exponential-argument}, and inserting the results into~\autoref{eq:approx-ratio-distribution} yields
\begin{equation}
    \label{eq:approx-distribution-almost-final}
    p_Z(Z) \propto \exp\Biggl(-\frac{\det\mathbf{C}^{-1}}{2} \frac{\bigl(\langle X \rangle - \langle Y \rangle Z \bigr)^2}{\mathbfit{q}^T \mathbf{C}^{-1} \mathbfit{q}}\Biggr).
\end{equation}
We can immediately see that the signal loss distribution is peaked around $Z = \langle X \rangle / \langle Y \rangle$, which is precisely the analytic expectation from the $m$-mode based calculation.

While the approximate distribution in~\autoref{eq:approx-distribution-almost-final} is not exactly Gaussian, the non-Gaussian tails are strongly suppressed.
We can take advantage of this and further simplify our result by approximating $\mathbfit{q} \approx \langle Y \rangle^{-1} \boldsymbol{\mu}$, which yields
\begin{equation}
    \label{eq:gaussian-signal-loss-distribution}
    p_Z(Z) \approx \frac{\langle Y \rangle^2}{\sqrt{2\pi\gamma\det\mathbf{C}}} \exp\Biggl( -\frac{\langle Y \rangle^2}{2} \frac{\bigl( \langle X \rangle - Z \langle Y \rangle\bigr)^2}{\gamma\det\mathbf{C}}\Biggr)
\end{equation}
after substituting back in the definition of $\gamma$, noting that $\det \mathbf{C}^{-1} = (\det \mathbf{C})^{-1}$, and adding an appropriate normalization.
So, for our case, $Z$ is approximately Gaussian distributed with mean $\langle X \rangle / \langle Y \rangle$ and variance $\gamma\det\mathbf{C} / \langle Y \rangle^4$.

\begin{figure*}
    \includegraphics[width=\textwidth]{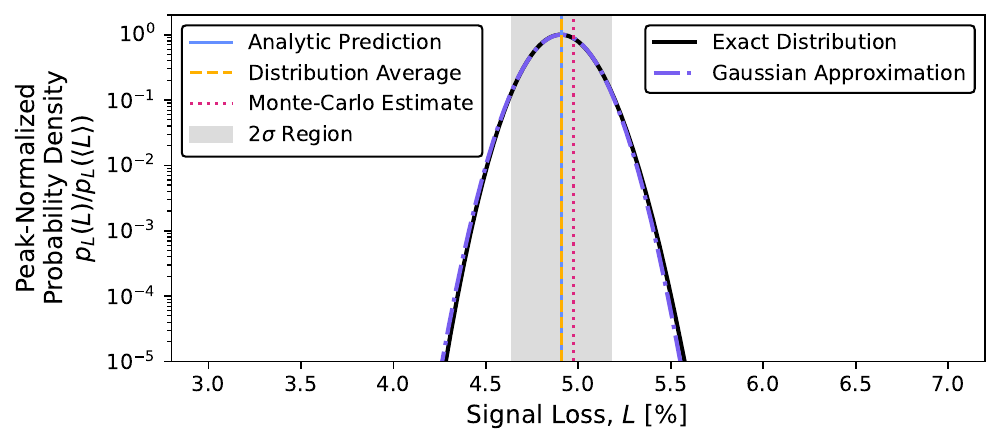}
    \caption{
        Predicted distribution of the signal loss estimated with a Monte Carlo, computed according to~\autoref{eq:approx-ratio-distribution}.
        The calculation was performed for a baseline with a 29-meter East-West projected length and a 25-meter North-South projected length, over a frequency range from 209 MHz to 220 MHz.
        The solid black line is the predicted distribution of signal loss estimates and the purple dot-dashed line is the Gaussian approximation of the distribution.
        The vertical lines show the expected signal loss computed various ways:
        The solid blue vertical line is the analytic prediction for the signal loss according to the ensemble average limit of~\autoref{eq:signal-loss-estimate-final};
        The dashed orange vertical line is the expected signal loss computed by numerically integrating the signal loss distribution;
        The dotted pink vertical line is the signal loss estimated from the visibility simulations used in Section~\ref{sec:SIMULATIONS}.
        The shaded grey area shows the $2\sigma$ confidence region.
        Note that the analytic prediction and the expected value from integrating the signal loss distribution appear to agree exactly, and the Monte Carlo estimated loss is well within the expected errors.
    }
    \label{fig:signal-loss-distribution}
\end{figure*}

We plot the exact peak-normalized distribution in~\autoref{fig:signal-loss-distribution} and compare it against the Gaussian approximation.
We use a baseline with a 29-m East-West projected length and a 25-m North-South projected length and consider frequencies between 209 MHz and 220 MHz for our sample calculation.
We additionally compare the expected signal loss obtained by numerically integrating the distribution against the analytic prediction from the $m$-mode formalism and the estimated signal loss from the suite of visibility simulations used in Section~\ref{sec:SIMULATIONS}.
We find that the Gaussian approximation is very accurate, but starts to slightly deviate from the exact distribution in the tails.
We compute the expected signal loss from the ratio distribution by numerically integrating $Zp_Z(Z)$ then converting the result to signal loss via $\langle L \rangle = 1 - \langle Z \rangle$.
We find that the expected signal loss from the analytic $m$-mode calculation is in excellent agreement with the expected signal loss from integrating the ratio distribution, and the estimated loss from the suite of visibility simulations agrees to within the expected error.

\subsection{Power Spectrum Covariance}
In this section, we compute the matrix elements of the covariance in~\autoref{eq:X-Y-covariance}.
In the calculations that follow, we assume that the cosmological signal is Gaussian distributed with unity variance at every frequency, and that the signal is uncorrelated between different frequencies (or equivalently, the cosmological signal has no spatial correlations).
Note that this is not the same as a cosmological signal with a flat power spectrum; however, the Blackman-Harris taper used in our power spectrum analysis heavily downweights values near the spectral window edges, so this analysis should be a good approximation of the simulation setup used in Section~\ref{sec:SIMULATIONS}.
Ultimately, these assumptions mean that 
\begin{equation}
    \bigl\langle V_{m\nu} V_{m'\nu'}^* \bigr\rangle = \delta_{mm'} \delta_{\nu\nu'} M_{m\nu},
\end{equation}
where $V_{m\nu}$ is the $m$-mode at frequency $\nu$.
It is possible to extend the analysis to consider a cosmological signal with more complicated statistics (which would likely require working with the beam transfer matrices $K_{\ell m}$ directly, as well as even moments of the cosmological field, instead of the $m$-mode power spectra), but such an analysis beyond the scope of this work.

The terms in the covariance are $a = \langle P_N'^2 \rangle - \langle P_N' \rangle^2, b = \langle P_N^2 \rangle - \langle P_N \rangle^2,$ and $c = \langle P_N P_N' \rangle - \langle P_N \rangle \langle P_N' \rangle$, where $P_N$ is the unfiltered power averaged over $N$ realizations of the sky and $P_N'$ is the filtered power averaged over $N$ realizations.
The average power can be computed via
\begin{equation}
    \label{eq:avg-power-over-N-realizations}
    P_N = \frac{u}{N_t N} \sum_{n=1}^N \sum_{t,\nu} B(\nu)^2 \sum_{m,m'} V_{m\nu}^r \bigl(V_{m'\nu}^r\bigr)^* e^{-i(m-m')\omega_\oplus t},
\end{equation}
where $B(\nu)$ is the frequency taper used in computing the delay spectrum, $N_t$ is the number of samples in the time series, $u = \bigl[ \sum_\nu B(\nu)^2 \bigr]^{-1}$, $r$ indexes different realizations (i.e., different trials in the Monte Carlo), and $V_{m\nu}$ are the $m$-modes at frequency $\nu$ (\autoref{eq:visibility-from-m-modes}).
The average power in the filtered data can be written as
\begin{equation}
    \label{eq:avg-filt-power}
    P_N' = \frac{u}{N} \sum_{r=1}^N \sum_\nu B(\nu)^2 \sum_{m,m'} X_{mm'} V_{m\nu}^r \bigl( V_{m'\nu}^r \bigr)^*,
\end{equation}
where
\begin{equation}
    \label{eq:m-mode-transfer-matrix}
    X_{mm'} = \frac{1}{N_t} \sum_t F_{tm} F_{tm'}^*
\end{equation}
and
\begin{equation}
    \label{eq:t-to-m-matrix}
    F_{tm} = \sum_{t'} W_{tt'} e^{-im\omega_\oplus t'}.
\end{equation}
$X_{mm'}$ can be thought of as the $m$-mode transfer matrix describing the action of the filter $W_{tt'}$ in the $m$-mode basis.
\autoref{eq:avg-power-over-N-realizations} and~\autoref{eq:avg-filt-power} are the two fundamental expressions we will use to compute the terms in~\autoref{eq:X-Y-covariance}.

The first moments of $P_N$ and $P_N'$ are straightforward to compute, since the product of $m$-modes is computed at a single frequency.
Skipping over the details, the expectation of $P_N$ can be computed via
\begin{equation}
    \label{eq:expected-average-power}
    \langle P_N \rangle = u \sum_\nu B(\nu)^2 \sum_m M_{m\nu},
\end{equation}
where $M_{m\nu}$ is the $m$-mode power spectrum (\autoref{eq:m-mode-power-spectrum}) at frequency $\nu$.
The calculation is also straightforward for the filtered power, which yields
\begin{equation}
    \label{eq:expected-filtered-power}
    \langle P_N' \rangle = u \sum_\nu B(\nu)^2 \sum_m X_{mm} M_{m\nu}.
\end{equation}
The second moments require a bit more work to compute, but the calculation is straightforward after realizing that
\begin{align}
    \Bigl\langle V_{m\nu}^n \bigl(V_{m'\nu}^n \bigr)^* V_{\mu\nu'}^{n'} \bigl(& V_{\mu'\nu'}^{n'}\bigr)^* \Bigr\rangle = \delta_{mm'} \delta_{\mu\mu'} M_{m\nu} M_{\mu\nu'}
    + \delta_{nn'} \delta_{m\mu'} \delta_{m'\mu} \delta_{\nu\nu'} M_{m\nu} M_{m'\nu},
    \label{eq:m-mode-fourth-moment}
\end{align}
which follows from assuming statistical independence of the cosmological signal at different frequencies.
Taking the expectation value of the square of~\autoref{eq:avg-power-over-N-realizations} and using~\autoref{eq:m-mode-fourth-moment} to simplify the result gives
\begin{equation}
    \langle P_N^2 \rangle = \langle P_N \rangle^2 + \frac{u^2}{N} \sum_\nu B(\nu)^4 \sum_{m,m'} M_{m\nu} K_{mm'} M_{m'\nu},
\end{equation}
where
\begin{equation}
    K_{mm'} = \Bigg| \frac{1}{N_t} \sum_t e^{-i(m-m')\omega_\oplus t} \Bigg|^2.
\end{equation}
So the variance in the average power is
\begin{equation}
    \label{eq:average-power-variance}
    \langle P_N^2 \rangle - \langle P_N \rangle^2 = \frac{u^2}{N} \sum_\nu B(\nu)^4 \sum_{m,m'} M_{m\nu} K_{mm'} M_{m'\nu}.
\end{equation}
This can be simplified somewhat by first performing the average over frequency and defining
\begin{equation}
    \mathcal{M}_{mm'} = u^2 \sum_\nu B(\nu)^4 M_{m\nu} M_{m'\nu},
\end{equation}
which allows us to rewrite the variance in the average power estimate as
\begin{equation}
    \langle P_N^2 \rangle - \langle P_N \rangle^2 = \frac{1}{N} \sum_{m,m'} K_{mm'} \mathcal{M}_{mm'}.
\end{equation}

The other terms in the covariance are relatively straightforward to compute with the help of~\autoref{eq:m-mode-fourth-moment}.
For the variance in the average power in the filtered visibilities, we find
\begin{equation}
    \langle P_N'^2 \rangle - \langle P_N' \rangle^2 = \frac{1}{N} \sum_{m,m'} X_{mm'} X_{m'm} \mathcal{M}_{mm'}.
\end{equation}
The covariance between the filtered and unfiltered average power estimates can be written as
\begin{equation}
    \langle P_N P_N' \rangle - \langle P_N \rangle \langle P_N' \rangle = \frac{1}{N} \sum_{m,m'} \Phi_{mm'} X_{m'm} \mathcal{M}_{mm'},
\end{equation}
where
\begin{equation}
    \Phi_{mm'} = \frac{1}{N_t} \sum_t e^{-i(m-m')\omega_\oplus t}.
\end{equation}
These expressions  provide enough auxiliary information to compute the distribution of signal loss estimates, provided the $m$-mode power spectrum and filter specification; however, care must be taken when implementing this numerically, since the arrays involved may become quite large.
As a compact reference, the results of this section can be summarized as
\begin{equation}
    \mathbf{C} = \frac{1}{N} \sum_{m,m'} \mathcal{M}_{mm'}
    \begin{pmatrix}
        |X_{mm'}|^2 & \Phi_{mm'} X_{m'm} \\
        \Phi_{mm'} X_{m'm} & |\Phi_{mm'}|^2
    \end{pmatrix},
\end{equation}
since $K_{mm'} = |\Phi_{mm'}|^2$ and $X_{mm'}^* = X_{m'm}$.
Taken together with~\autoref{eq:avg-power-over-N-realizations} and~\autoref{eq:avg-filt-power}, this means that the results of the Monte Carlo analysis presented in Section~\ref{sec:SIMULATIONS} will converge to the predictions of our $m$-mode-based formalism, with the error in the Monte Carlo estimates decreasing with the square root of the number of visibility simulations.


\end{document}